\documentclass[graybox, secnum]{svmult}

\usepackage{mathptmx}       
\usepackage{helvet}         
\usepackage{courier}        
\usepackage{type1cm}        
\usepackage{makeidx}         
\usepackage{graphicx}        
\usepackage{multicol}        
\usepackage[bottom]{footmisc}
\usepackage{hyperref}        
\usepackage{soul}            
\usepackage{amssymb}            
\hypersetup{colorlinks=true,urlcolor=blue}
\usepackage[square,numbers]{natbib}
\usepackage{xcolor, soul}
\sethlcolor{green}
\makeatletter
\newcommand{\dummylabel}[2]{\def\@currentlabel{#2}\label{#1}}
\makeatother
\makeindex             

\begin{document}

\dummylabel{ch:acc_disk_theory}{1} 
\dummylabel{ch:binary_evol}{2} 
\dummylabel{ch:white_dwarfs}{3} 
\dummylabel{ch:accretion}{4} 
\dummylabel{ch:ejection}{5} 

\title*{Accreting white dwarfs}
\author{Natalie Webb}
\institute{Natalie Webb \at Institute de Recherche en Astrophysique et Plan\'etologie, 9 avenue du Colonel Roche, 31028 Toulouse, France \email{Natalie.Webb@irap.omp.eu}
}
%
%
\maketitle
\abstract{White dwarfs are the most common endpoints of stellar evolution. They are often found in close binary systems in which the white dwarf is accreting matter from a companion star, either via an accretion disc or channelled along the white dwarf magnetic field lines. The nature of this binary depends on the masses and the separation of the two stellar components, as well as other parameters such as the white dwarf magnetic field and the nature of the stars.  This chapter looks at the formation of white dwarfs and intrinsic properties, before looking at the different populations of accreting white dwarf binary systems that exist. The chapter covers the characteristics of the various sub-populations and looks at how they evolve. The means to discover and study various sub-classes of
accreting white dwarfs in the optical and other bands is discussed, and the role of
these systems in the broader astrophysical context is considered. Future missions that will find new systems and new populations are also reviewed. Finally some of the current open questions regarding accreting white dwarfs are presented.}

\section{Keywords} 
Close binaries -- Novae -- Cataclysmic variables -- White dwarfs -- Accretion discs -- Surveys -- Globular clusters -- Supernovae Ia

\section{Introduction}

\hl{White dwarfs} are formed at the end of the lives of the least massive stars and are thus the most common endpoints of stellar evolution. White dwarfs were the first type of compact object to be discovered. William Herschel made the first detection of a white dwarf, 40 Eridani B in 1783 \cite{hers85}. However, it wasn't until 1915 when \cite{adam15} demonstrated that Sirius' companion had a mass of around a solar mass. However, it had a low luminosity similar to that of 40 Eridani B \cite{adam14}, revealing that these stars were different from the general population of stars observed. The approximately flat optical spectrum ($\sim$4000-7000\AA), along with the low luminosity, which indicates a small radius, caused \cite{luyt22} to refer to these stars as white dwarfs.  By 1924 it was clear that white dwarfs were very different to other stars known \cite{eddi24}, but it was only in 1931 that \cite{chan31} understood the nature of the dense degenerate matter constituting these objects. Today, around a hundred thousand white dwarfs are known. Many are found in the Solar neighbourhood \cite[e.g.][]{jime18} and some are young enough and therefore bright enough to be visible through a pair of binoculars (e.g. 40 Eridani B, magnitude $\sim$ 9.5). 

White dwarfs are characterised by small radii, typically similar to the radius of the Earth, but with masses reaching 1.4 M$_\odot$ \cite{chan31}. This implies high densities and therefore strong gravitational fields, which become stronger as the mass increases (and the radius decreases, see Section~\ref{sec:chars_WD}), so that it becomes necessary to use general relativity to understand their nature and their environment. If a white dwarf is in a close \hl{binary} system, matter from the companion star may be lost through the inner Lagrangian point of the binary (see Chapter~\ref{ch:acc_disk_theory}) and become trapped in the gravitational field of the white dwarf, allowing it to fall onto the white dwarf, rendering it hotter and brighter than an isolated white dwarf of a similar age. Studying the evolution of these systems can give an insight into the white dwarf nature and help to understand the accretion phenomenon which is found throughout the Universe, in stellar and planetary formation, as well as around supermassive black holes in the hearts of galaxies.

This chapter covers white dwarfs and the nature of the different binary systems containing these compact objects, as well as the different populations of these systems, where and how they are found and how they can be used to understand different phenomena in the Universe.

\section{What is a white dwarf?}
\label{sec:whatis_wd}

In main-sequence stars, i.e. stars burning hydrogen in their core, the hydrostatic equilibrium is achieved thanks to the gas pressure opposing the pressure due to the mass of the star. However, white dwarfs are no longer undergoing fusion.  For white dwarfs to find a stable equilibrium, a different pressure is required and this is through electron \hl{degeneracy}. The physics of the electron degeneracy dictate the white dwarf characteristics and properties, including the maximum mass allowed and how the white dwarf responds to accreted matter. The electron degeneracy and its implications for the nature of the white dwarf are described in the following subsections.

\subsection{Electron degeneracy}
\label{sec:WD_degeneracy}

The pressure, $P_e$, that electrons ($e^-$) can exert is a function of their density ($\rho_e$) i.e. $P_e$ = $f(\rho_e$). To determine this pressure, the number of free-electron states in a volume, $V$, are calculated. For simplicity, a cube with the length of the side, $L$, is considered. This means $V=L^3$.  Duplicating this cube to fill all the space available and allowing the wave vectors of the free-electron quantum states to take only discrete values, the electron wave function is given by $\psi \propto e^{(ik·x)}$, when considering the three dimensions, $x, y, z$, $k$ = ($k_x$,$k_y$,$k_z$), so that the replication of the cube leads to:

\begin{equation}
k_x =  n_x\frac{2\pi}{L}  \quad {\rm where} \quad n_x = 1, 2, \ldots
\end{equation}

The allowed states are separated by $2\pi/L$, so the density of the states in $k$ space ($N$) is 

\begin{equation}
dN =  g\frac{L^3}{(2\pi)^3} d^3k  \hspace*{2cm} (d^3k \equiv dk_xdk_ydk_z)
\end{equation}
where $g$ is the degeneracy factor for spin (spin can be either up ($\uparrow$) or down ($\downarrow$)).

Taking the de Broglie relationship for the electron's momentum, $p$,  associated with its wave vector: $p = \hslash\ k$, where $\hslash$ is the Hslash constant divided by 2$\pi$, the density of the states in $k$ space is converted to momentum space using:

\begin{equation}
dN =  g\frac{L^3}{(2\pi\hslash)^3} d^3p  
\end{equation}

The number density of particles per unit volume ($n$) with momentum states in $d^3p$ is then

\begin{equation}
dn =  g\frac{1}{(2\pi\hslash)^3} f(p)d^3p  
\label{eq:particle_no_density}
\end{equation}
where $f(p)$ is the mode occupation  number, i.e. the number of particles in the cube with a particular wave function. For bosons (e.g. photons), $f(p)$ is not restricted, but fermions (i.e. electrons with spin angular momentum $\hslash$/2) must obey the Pauli exclusion principle. This states that

\begin{equation}
f(p) \leq 1 
\label{eq:pauli}
\end{equation}

The electron gas is thus different to a classical gas and the Pauli exclusion principle limits the density of the electron gas.

\subsection{The equation of state of electron-degenerate matter}

Treating the particle momenta as a Maxwellian distribution, where each velocity ($v$) component has a Gaussian distribution with a standard deviation, gives the following equation

\begin{equation}
\Psi(v) = \frac{1}{(2\pi\sigma^2)^{3/2}}e^{(-v^2/2\sigma^2)}d^3v 
\label{eq:Maxwell_dist}
\end{equation}

The velocity dispersion ($\sigma$) is a function of the temperature, $T$, through the equipartition of energy: $\sigma^2=k_BT/m$ (where $k_B$ is the Boltzmann constant and $m$ the particle mass). Writing Eq.~\ref{eq:Maxwell_dist} in terms of the number density of particles in a given momentum range and then multiplying by the total number density of particles and using the momentum equation $p=mv$ gives,

\begin{equation}
dn = \frac{n}{(2\pi mk_BT)^{3/2}}e^{(-p^2/2mk_BT)}d^3p 
\end{equation}

Comparing Eqs.~\ref{eq:Maxwell_dist} and \ref{eq:particle_no_density} implies a critical density ($n_{crit}$) where the classical description would violate the Pauli exclusion principle (Eq.~\ref{eq:pauli}) for $p = 0$.

\begin{equation}
n_{crit} = \frac{g}{(2\pi )^{3/2}}\frac{(mk_BT)^{3/2}}{\hslash^3} 
\end{equation}

The gas will then be in the classical regime at high temperatures and fixed density, but quantum effects become important as $T\rightarrow$\ 0. Integrating Eq.~\ref{eq:particle_no_density} over all momentum states leads to the total number density of particles:

\begin{equation}
n = \frac{g}{(2\pi \hslash )^{3}}\int{f(p) d^3p} 
\end{equation}

Then as $T\rightarrow$\ 0, the states are occupied only up to the Fermi momentum ($p_F$).  Hence:

\begin{equation}
n = \frac{g}{(2\pi \hslash )^{3}}\int_0^{p_F}{d^3p = \frac{g}{(2\pi \hslash )^{3}}\frac{4\pi}{3}p^3_F} 
\label{eq:Tzero}
\end{equation}

The Fermi momentum is then correlated with the particle density as shown in Eq.~\ref{eq:particle_density}, 

\begin{equation}
p_F = 2\pi \hslash \left ( \frac{3}{4\pi g} \right ) ^{1/3} n^{1/3} 
\label{eq:particle_density}
\end{equation}

If the gas density increases, the Fermi momentum will also increase. The additional particles are forced to fill the higher momentum states because the lower momentum states have already been occupied.

The pressure from the electrons can be determined by considering the pressure as a momentum flux and therefore the flux of electrons in the $x$-direction is just the number of electrons crossing a unit area per unit time, or $n_e v_x$.  The pressure is then approximated by $P \sim p_x n_e v_x$. Using Eq.~\ref{eq:particle_no_density} the pressure in the $x$-direction (which by isotropy must be equal to the pressure in any direction) is 
\begin{equation}
P = P_x = \frac{g}{(2 \pi \hslash)^3} \int p_x v_x f(p) d^3p 
\end{equation}

Using spherical polar coordinates in momentum space

\begin{equation}
\int{p_x v_x dp_x dp_y dp_z} = \frac{1}{3} \int{(p_xv_x + p_yv_y + p_zv_z)dp_x dp_y dp_z} =  \frac{1}{3} \int{p·v4\pi p^2dp} 
\end{equation}
so 
\begin{equation}
P = \frac{g}{3}\frac{1}{(2\pi \hslash)^3}\int_0^\infty p v f(p)4\pi p^2 dp 
\end{equation}

As $T \rightarrow$\ 0 (Eq.~\ref{eq:Tzero}), electron speeds do not achieve relativistic values so that $pv = p^2/m_e$, and

\begin{equation}
P_e = \frac{4 \pi g}{3(2 \pi \hslash)^3}\int_0^{p_F} \frac{p^2}{m_e} p^2 dp = \frac{g}{30\pi^2 \hslash^3 m_e}p_F^5 
\label{eq:electron_pressure}
\end{equation}

Rewriting the pressure as a function of density $\rho_e = m_e n_e$ using Eq.~\ref{eq:particle_density} 

\begin{equation}
p_F = \left ( \frac{6 \pi \hslash \rho_e}{g m_e} \right ) ^{1/3} 
\end{equation}
and substituting into Eq.~\ref{eq:electron_pressure}, the pressure for non-relativistic electrons can be determined as follows:

\begin{equation}
P_e = \frac{g}{30 \pi^2 \hslash^3}  \left ( \frac{6 \pi^2 \hslash^3 }{g} \right )^{5/3} \rho^{5/3} m_e^{-8/3} = K_1 \rho_e^{5/3}  \hspace{0.3cm}
\label{eq:Pe_non_rel}
\end{equation}
\begin{equation}
{\rm where}\ \ K_1 = \frac{\pi^2 \hslash^2}{5 m_e^{8/3}} \left ( \frac{6}{g \pi} \right )^{2/3} 
\end{equation}

As before, as the densities increase (Eq.~\ref{eq:particle_density}), the  Fermi momentum increases until it reaches relativistic values. Some particles may be compelled into momentum states with velocities approaching the speed of light, $c$. Replacing $v$ with $c$ gives the relativistic expression

\begin{equation}
P_e = \frac{4 \pi g}{3(2 \pi \hslash)^3}\int_0^{p_F} p c  p^2 dp = \frac{g\ c}{24\pi^2 \hslash^3}p_F^4 
\label{eq:Pe_rel}
\end{equation}
or 

\begin{equation}
P_e = \frac{gc}{24 \pi^2 \hslash^3}  \left ( \frac{6 \pi \hslash^3 \rho_e}{g m_e} \right )^{4/3}  = K_2 \rho_e^{4/3} \hspace{0.3cm}
\label{eq:Pe_rel2}
\end{equation}
\begin{equation}
{\rm where}\ \ K_2 = \frac{\pi \hslash c}{4 m_e^{4/3}} \left ( \frac{6}{g \pi} \right )^{1/3} 
 \end{equation}

Equations \ref{eq:Pe_non_rel} and \ref{eq:Pe_rel2} reveal that in a degenerate gas, the pressure depends on the density alone and is thus independent of temperature. In a partially  degenerate  gas a residual temperature dependence remains.  

Only the electrons are considered in determining the pressure. In Eqs.~\ref{eq:Pe_non_rel} and \ref{eq:Pe_rel2} it can be seen that the particle mass is in the denominator. As the proton is 1836 times more massive than the electron, its contribution to the pressure is therefore negligible. 

\subsection{The Chandrasekhar mass}
\label{sec:chandra_mass}

As the white dwarf is supported by electron degeneracy, there must be a limit to the pressure that can be provided, simply because there is a limit to the phase space density of the electrons. Therefore, it is intuitive that there is a maximum mass for the white dwarf. This mass can be deduced by estimating the energy density, $U$, of the degenerate gas in a similar way to that which was done for the pressure. Integrating over momentum space and incorporating a term, $\epsilon(p)$, to convey the energy per mode, the energy density is given by

\begin{equation}
U = \frac{g}{(2\pi \hslash)^3}\int_0^\infty \epsilon(p) f(p) 4\pi p^2 dp 
\label{eq:energy_density}
\end{equation}

As $T\rightarrow$\ 0, $f(p) = 1$, but is zero otherwise. In the relativistic case $\epsilon(p) = pc$. Using this in Eq.~\ref{eq:energy_density} and integrating gives rise to the electron energy density ($U_e$)

\begin{equation}
U_e = \frac{g}{(2\pi \hslash)^3}\frac{1}{4}4\pi c p_F^4 
\label{eq:Urel}
\end{equation}

Using Eq.~\ref{eq:particle_density} to express $U_e$ as a function of the electron number density gives

\begin{equation}
U_e = \frac{3}{4} \left ( \frac{6 \pi^2}{g} \right )^{1/3} \hslash c n_e^{4/3}
\label{eq:Ue_number_density}
\end{equation}

The non-relativistic case for $\epsilon(p) = p^2/2m_e$ then gives 

\begin{equation}
U_e = \frac{3\hslash^2}{10m_e} \left ( \frac{6 \pi^2}{g} \right )^{2/3}  n_e^{5/3}.
\label{eq:Ue_non_rel_number_density}
\end{equation}

The total kinetic energy ($E_K$) of the degenerate  electrons is  proportional to their energy density multiplied by the volume.  In the relativistic case, $E_K \propto U_e V \propto n_e^{4/3}V \propto M^{4/3}/R$, where $M$ is the mass and $R$ is the radius of the white dwarf. The gravitational energy ($E_p$) is proportional to $M^2/R$. The total energy ($E_{tot}$) is then the sum of both terms : $E_{tot} = (AM^{4/3} - BM^2)/R$, where $A$ and $B$ are constants. For a set of masses, the total energy must be positive. The total energy is reduced as the radius increases (the star expands) at which point the electrons become only mildly relativistic. The star then exists as a stable white dwarf. If however the mass increases, the total energy becomes negative and the radius decreases until gravitational collapse is inevitable, see Section~\ref{sec:type1a} for more discussion on this point.

To determine the mass at which the electron degeneracy pressure is no longer sufficient to support the white dwarf, $A$ and $B$ must be determined \cite{chan31}. These values depend on the molecular weight per electron of the material inside the white dwarf and the density profile. The critical mass at which the electron pressure is overcome was first determined by Subrahmanyan Chandrasekhar \cite{chan31}, and is named after him, the \hl{\it Chandrasekhar mass}. This was subsequently determined to be $\sim$1.4 M$_\odot$ \cite{ostr66}.

\subsection{White dwarf formation}
\label{sec:WD_form}

Stars with masses up to $\sim$8 - 10.6 M$_\odot$, depending on the initial stellar metallicity \cite{maed89,meyn91,ibel13,dohe15,woos15}, leave behind white dwarfs at the end of their lives.  There are different types of white dwarf (see Section~\ref{sec:chars_WD}), and their nature is often related to the initial mass of the progenitor. Stars on the main sequence of the Hertzsprung Russell diagram undergo hydrogen fusion in their core, providing a force that counteracts the gravitational force due to the mass of the star, so that the star remains in hydrostatic equilibrium. Hydrogen fusion produces helium. When insufficient hydrogen remains in the core to continue the fusion process, the star leaves the main sequence with a helium-flash. For the lowest mass stars, there is insufficient pressure and the temperature is too low for the helium to undergo fusion. These stars leave behind remnants that are constituted mainly of hydrogen or helium. Stars with a mass $\gtrsim$1.85 M$_\odot$ \cite{meyn91} can continue to burn helium in a stable fashion and produce carbon-oxygen cores. Once insufficient matter remains for the fusion process to continue, the gravitational force due to the mass of the star dominates. The outer parts of the star are blown away during the planetary nebula phase and the core will then collapse due to gravity.

\subsection{White dwarf characteristics}
\label{sec:chars_WD}

\subsubsection{Rotation rates}

In non-degenerate, i.e. {\it normal} matter, the volume increases with the mass. For degenerate matter, the opposite occurs, so as the mass of the white dwarf increases, the radius decreases (see also Sec.~\ref{sec:chandra_mass}).  This was described by \cite{naue72},

\begin{equation}
R = 7.8 \times 10^{8} {\rm cm} \left [ \left ( \frac{1.44}{M} \right ) ^{2/3} - \left ( \frac{M}{1.44} \right ) ^{2/3}  \right ]^{1/2}
\label{eq:WDmass}
\end{equation}

The radius of a solar mass white dwarf is approximately the same as the radius of the Earth ($\sim$6000 km). Less massive white dwarfs have larger radii, for example a 0.7 M$_\odot$ white dwarf would have a radius of $\sim$7500 km. Chandrasekhar mass white dwarfs have smaller radii, $\sim$1200 km. The compact nature of white dwarfs implies that the gravitational force responsible for holding it together ($GMm/R^2$) where $m$ is the mass of a particle on the surface of the white dwarf, is extremely high. This means that white dwarfs can rotate extremely rapidly without flying apart due to the centripetal force ($mv^2/R$).  The velocity ($v$) at the surface of the star is $(2 \pi r)/P$, where $P$ is the period of the rotation of the star and therefore the theoretical minimum rotational period of a white dwarf can be calculated

\begin{equation}
\frac{G M m}{R^2} > \frac{m 4\pi^2 R}{P^2} \hspace{0.5cm} \Rightarrow \hspace{0.5cm} P > 2 \pi \sqrt{\frac{R^3}{GM}}
\label{eq:min_rot}
\end{equation}

From Eq.~\ref{eq:min_rot}, the minimum period is for a minimum radius, which corresponds to a white dwarf with the Chandrasekhar mass. This implies a minimum rotation period of $\sim$0.5 s. Without energy injected into the system to maintain this high rotation rate, the white dwarf will rapidly spin down. Indeed, no white dwarf with such a short period has been detected to date, where the shortest rotation period of a white dwarf is 24.93 s for the white dwarf in the cataclysmic variable LAMOST J024048.51+195226.9 \cite{peli21}, as mass transfer in such a binary can transfer angular momentum to the white dwarf, spinning it up, see Section~\ref{sec:dn_nl}.

\subsubsection{Magnetic field}

As angular momentum should be conserved as the star evolves to a white dwarf, so the magnetic flux may also be conserved. Some main-sequence stars can show magnetic fields as high as 3 $\times$ 10$^2$ - 3 $\times$ 10$^4$ G\footnote{1 G (Gauss) is the CGS unit for magnetic fields and corresponds to $10^{-4}$ T (Tesla). } \cite{wick10}. The magnetic flux, $B$, is proportional to the inverse of the radius squared and therefore magnetic fields of the order 10$^8$ G can be generated when the main sequence star collapses to the white dwarf \cite{wick10}. White dwarfs with lower magnetic fields of 10$^3$-10$^4$ G may have had their fields generated by convective dynamos during stellar evolution to the white dwarf state or for fields of 10$^5$-10$^6$ G, through contemporary dynamos in cooling white dwarfs that develop extensive convective envelopes.

\subsubsection{Temperature and cooling}

White dwarfs are born with high temperatures. The initial temperature can be estimated using the contraction of the thermally unsupported stellar core down to the radius at which degeneracy pressure will halt the contraction.  The virial theorem states that just before reaching the final point of equilibrium,  the thermal energy ($E_{th} = \frac{3}{2}Nk_BT$) will equal half of the potential energy ($GM^2/R$). For a pure helium composition, the number of nuclei in the core is $M/4m_p$, and the number of electrons is $M/2m_p$, so

\begin{equation}
E_{th} = \frac{3}{2}\frac{M}{m_p}\left (  \frac{1}{2} + \frac{1}{4} \right ) k_BT =  \frac{9}{8}\frac{M}{m_p}k_BT 
\label{eq:thermal_energy}
\end{equation}

Using the virial theorem,

\begin{equation}
k_BT \sim \frac{4}{9}\frac{GMm_p}{R}      
\end{equation}

for a solar mass white dwarf, its temperature at birth is $\sim1 \times 10^9$ K. This hot temperature implies that the emission peaks in the X-ray range. Once the core becomes an exposed white dwarf, its radiation ionises the gas blown off during the Asymptotic Giant Branch (AGB) phase, giving rise to the planetary nebula discussed above. The white dwarf will start to cool. Neglecting the envelope around the white dwarf and assuming a uniform temperature throughout, the maximum cooling rate can be determined.  The radiative energy loss is obtained by equating the luminosity, $L$, of the blackbody, given by the Stefan-Boltzmann law, to the rate of change of thermal energy (Eq.~\ref{eq:thermal_energy})

\begin{equation}
L = 4 \pi R^2 \sigma T^4 \sim \frac{dE_{th}}{dt} = \frac{3}{8}\frac{Mk_B}{m_p}\frac{dT}{dt}
\end{equation}
where $\sigma$ is the Stefan-Boltzmann constant. Separating the variables $T$ and $t$ and integrating, the cooling time to reach a temperature $T$ is

\begin{equation}
\tau_{cool} \sim \frac{3}{8}\frac{Mk_B}{m_p}\frac{1}{4 \pi R^2 \sigma}\frac{1}{3}\frac{1}{T^3} \sim 3 \times 10^9 {\rm yr} \left ( \frac{T}{10^3 K} \right )^{-3}
\end{equation}

In reality this is too rapid. The outer, non-degenerate envelope acts as an insulator and the inner regions crystallise, modifying the cooling with time \cite[e.g.][]{wood92}. One of the interests in understanding exactly how white dwarfs cool is that they have cooling times comparable to the age of the Universe. It may therefore be possible to use these objects to determine the history of star formation in our Galaxy by modelling their luminosity function. Studying these systems in binary systems in which accretion onto the surface of the white dwarf occurs (see Section~\ref{sec:binaries}), can help understand how white dwarfs cool as the white dwarf is constantly being heated and cooled due to increasing and decreasing accretion.

\subsubsection{Composition}

Most white dwarfs are composed of carbon and oxygen, but the atmosphere is usually observed to be predominantly hydrogen ($\sim$83\% of spectroscopically confirmed white dwarfs). White dwarfs with hydrogen atmospheres are called DA white dwarfs.  7\% of white dwarfs \cite[e.g.][]{ouri19} show helium spectra and are known as DO white dwarfs if they are hot enough to show He II lines (from singly ionised helium) and DB if they only show He I lines (from neutral helium). DC white dwarfs are too cool to show absorption lines in their optical spectra ($\sim$5\% of white dwarfs) and are believed to have helium atmospheres which explains why hydrogen lines are not seen \cite[e.g.][]{ouri19}. DQ white dwarfs, showing carbon lines or bands, count for only 1\% of white dwarfs. White dwarfs with other types of metal lines in their optical spectra are known as DZ white dwarfs and make up about 4\% of the population \cite[e.g.][]{ouri19}. The atmosphere is always dominated by the lightest element present, as the high surface gravity leads to a stratification of the atmosphere. 

Due to the high density of white dwarfs, the atmosphere also has a high pressure. In the optical and ultra-violet spectra, broad line wings are observed due to pressure broadening \cite[or collisional broadening,][]{fole46}, as collisions between atmospheric atoms reduce the effective lifetime of a state, leading to broader lines. This is most notable in the Balmer lines seen in the optical spectra of DA white dwarfs, but as white dwarfs are hot and therefore also emit strongly at shorter wavelengths, the effect is also seen in the Lyman lines in the ultra-violet, for example \cite{bars03}. Isolated white dwarfs can be detected in soft (low energy) X-rays, but the spectra are thermal in nature and do not usually show any lines. X-ray emission is stronger when the white dwarf is in a binary system accreting from a companion star (see Section~\ref{sec:binaries} and Chapter~\ref{ch:white_dwarfs}).

\subsubsection{Observed masses and radii}

The {\it Montreal White Dwarf Database} \cite{dufo17} provides data for $\sim$56000 spectroscopically confirmed white dwarfs. The average temperature for the white dwarfs in this database is $\sim$7500 K, but with a large range spanning $\sim$3000-180000 K.  The mean mass of the sample is 0.57 M$_\odot$, but with the range spanning very low ($\sim$0.01 M$_\odot$) masses up to masses in slight excess of the Chandrasekhar mass, $\sim$1.49 M$_\odot$. The average surface gravity is log(g) $\sim$ 8, with the range spanning 3.1-10. The magnetic fields are in the range 1 $\times$ 10$^5$ -  9 $\times$ 10$^8$ G. These values are similar to the almost complete sample of white dwarfs detected out to 100 pc with {\it Gaia} \cite{jime18}.  This survey also gives an average white dwarf radius of $\sim$ 0.012 R$_\odot$ (8346 km), with a range 0.002-0.02 R$_\odot$ (1391-13910 km), but with a possible bimodal distribution with maxima at 0.01 R$_\odot$ (6955 km) and at 0.0125 R$_\odot$ (8694 km), which remains to be explained.

\section{Accreting white dwarfs}
\label{sec:binaries}

The majority of stars are found in binary systems, where the two stars orbit about their common centre of mass. The \hl{orbital period}, or the time it takes for the two stars to orbit each other can be thousands of years ($P_{orb}$). These systems will evolve as described in the Chapter~\ref{ch:binary_evol} and one or both of the stars can evolve to become a white dwarf, depending on its initial mass, see Section~\ref{sec:WD_form}. Due to the compact nature of the white dwarf (see Section~\ref{sec:WD_degeneracy}), the orbital period can become very short (see Section~\ref{sec:periods}), as little as tens of minutes, so that the stars are very close, and they will have a strong influence on one another. The different types of close binary systems containing white dwarfs are discussed here.

\subsection{Roche Lobe Overflow and accretion}
\label{sec:acc}

A full account of accretion and ejection is beyond the scope of this Chapter which are discussed in Chapters~\ref{ch:accretion} and \ref{ch:ejection}, however, several notions concerning these topics are necessary to understand some of the aspects covered in this Chapter.

To understand a binary system and the mutual influence of the two stars, the total potential (gravitational and centripetal forces), should be considered. This is approximated by the Roche potential, $\Phi_R$ and assumes point like stars in circular orbits. Considering Cartesian coordinates ($x,y,z$) that rotate with the binary, the $x$-axis is described by the line traversing the stellar centres  (separated by a distance $a$ and of masses $M_1$ and $M_2$), the $y$-axis is in the direction of orbital motion of the primary and the $z$-axis is perpendicular to the orbital plane. The gravitational potential $\Phi_G$ is then defined as

\begin{equation}
\label{eq:RocheGravPotential}
\Phi_G = -\frac{GM_1}{(x^2 + y^2 + z^2)^{0.5}}-\frac{GM_2}{[(x-a)^2 + y^2 + z^2]^{0.5}}
\end{equation}
and the centripetal potential $\Phi_C$ is written as 

\begin{equation}
\label{eq:RocheCentripetalPotential}
\Phi_C = -\frac{1}{2} \Omega^2[(x-\mu a)^2 + y^2]
\end{equation}
where $\mu$ is the mass of the second star ($M_2$) with respect to their combined masses ($M_1 + M_2$), also known as the mass function and $\Omega$ is $2 \pi/P_{orb}$. The Roche potential can be written as 

\begin{equation}
\label{eq:RocheCentripetalPotential2}
\Phi_R = -\frac{GM_1}{(x^2 + y^2 + z^2)^{0.5}}-\frac{GM_2}{[(x-a)^2 + y^2 + z^2]^{0.5}} -\frac{1}{2} \Omega^2[(x-\mu a)^2 + y^2]
\end{equation}

Close to the star, the potential is dominated by its gravitational potential, so the stellar surfaces are almost spherical. Further from the stellar centre, both effects become important: the tidal effect, which causes an elongation in the direction of the companion star and flattening due to the centripetal force. The surfaces are then distorted so that their major axis is along the line connecting the stellar centres. The innermost equi-potential surface encompassing both stars, defines the {\it Roche lobe} of each star. Matter inside a Roche lobe is gravitationally bound to that star. 

The topology of the surfaces can be determined by the Lagrange points, the point at which the gradient of the potential becomes zero. There are five such points where the effective gravity from the two stars and the centrifugal force cancel. The inner Lagrangian point, $L_1$ (also called the `saddle point'), forms a pass between the two Roche lobes. $L_1$, $L_2$ (situated behind the Roche lobe corresponding to M$_2$) and $L_3$ (behind the Roche lobe corresponding to M$_1$) are unstable, i.e. a small perturbation will displace a body placed at these points. However, $L_4$ (situated above L$_1$ in the rotation plane) and $L_5$ (situated below L$_1$ in the rotation plane)  are stable, i.e. resist gravitational perturbations.

If one star fills its Roche lobe exactly, e.g. M$_2$, while the other, here the white dwarf, is still smaller than its Roche lobe, the binary is said to be semi-detached, or an interacting binary.   If the star $M_2$ expands, it will encounter a {\it hole} in its surface, near $L_1$. Here, hydrostatic equilibrium is no longer possible and matter will flow through the {\it nozzle} around $L_1$ into the Roche lobe of the white dwarf. This is known as Roche-lobe overflow. The matter that arrives in the white dwarf Roche lobe is gravitationally bound to it. The matter has angular momentum, due to the rotation of the binary system and so it will circle around the white dwarf, eventually forming a ring. Due to friction in the ring, the material will spread into a disc shape, known as an {\it accretion disc}.

The matter forms a thin disc around the white dwarf if the mass transfer rate is low. In this case, the height of the disc is much smaller than the radius (typically 0.1-3\%). The matter loses angular momentum due to frictional forces within the disc, allowing matter to fall onto the white dwarf, heating it up and causing it to become brighter, either over the whole surface or at the poles, depending on the magnetic field configuration, see Section~\ref{sec:polsandips}. The accretion disc is cool towards the exterior, mainly emitting in the infra-red, but becomes hotter towards the centre where it achieves high enough temperatures that result in X-ray emission. Each disc annulus emits a blackbody spectrum that corresponds to the temperature of that annulus. The accretion disc then emits the totality of the annuli, called a disc blackbody. The lowest frequencies $\nu$ are the Rayleigh-Jeans tail emission from the outer regions of the disc, where the intensity $I_\nu \propto \nu^2$. The middle part of the spectrum results from the sum of the blackbodies ($I_\nu \propto \nu^{1/3}$) and the highest frequencies emanate from the inner edge of the accretion disc with a temperature $T_{in}$, that has an exponential cutoff, $I_\nu \propto e^{-h\nu/kT_{in}}$. Spectral lines may be superposed on this emission. 

Accretion is an efficient way of producing energy.  As particles with mass $m$ fall from infinity onto the compact white dwarf, they lose their gravitational potential energy and the change in energy $\Delta E_{acc}$ is given by 

\begin{equation}
\Delta E_{acc} =\frac{G M m}{R}
\label{eq:acc}
\end{equation}
where $M$ and $R$ are the mass and the radius of the white dwarf.

Outbursts occur as matter builds up in the disc, so that the density and therefore the temperature increases. At a critical density, the matter, essentially hydrogen if accreting from a hydrogen rich star, ionises, and thus becomes more viscous, allowing matter to pass through the disc and fall onto the white dwarf. This results in an important increase in the luminosity until sufficient matter is evacuated from the disc, so that it becomes less dense and cooler, thus ending the outburst. Theory predicts that the outbursts start either in the outer regions of the disc and move inwards through the disc, and are termed as {\it outside-in} or they start in the inner regions of the disc and move outwards, and are called {\it inside-out} outbursts \cite{smak84}. It
is unclear prior to an outburst whether the outburst is going to be
outside-in or inside-out.  It is likely that the type of outburst that
takes place, is due to the amount and location of material left in the
disc after the previous outburst \cite{smak84}. For a detailed treatment see Chapter~\ref{ch:acc_disk_theory}.  

\subsection{Outflows and jets}

Momentum can be extracted from the radiation field of the accretion disc through line opacity \cite{cast75}, driving a wind (or outflow) from the accretion disc surface. However, magnetic fields threading the accretion disc can also play a role in launching the wind. Broad, blue-shifted, high excitation absorption lines, along with P Cygni line profiles observed from cataclysmic variables (novalikes and dwarf novae in outburst) seen face on are evidence for winds from the accretion discs in these systems.  These winds are not thought to be completely radiatively driven, as wind line strengths are not directly correlated with the strength of the photo-ionising continuum e.g. \cite{fron05}. 

Jets, collimated beams of high velocity material, often emitted perpendicular to the accretion disc have been detected in some accreting white dwarf systems. The energy to launch jets is thought to come either from the rotation of the compact object \cite{blan77} or from differential rotation in the accretion disc \cite{blan82}. Therefore, it is logical to expect that jets may be launched from accreting white dwarfs if the magnetic field is not so strong that it disrupts the central regions of the disc. Indeed, jets have been detected from a wide range of accreting white dwarfs, including supersoft sources, e.g. \cite{tomo98} and in symbiotic systems e.g. \cite{tayl86}. Alternative models also exist. Observational evidence shows that the jets are often brightest in the radio domain, with spectra indicating that the emission mechanism is synchrotron, supporting these theories.

Evidence for jet emission from a cataclysmic variable, outside a nova eruption, was first detected from SS Cyg \cite{koer08}. These jets appear to be similar to jets detected in X-ray binaries (stellar mass black holes or neutron stars accreting from a companion star), as radio flares best explained by transient jets were detected as SS Cyg moved from quiescence to outburst in the same way as X-ray binaries. The power associated with these jets was analogous to X-ray binary jets. However, \cite{hame17} state that the evolution throughout the outburst observed in cataclysmic variables and that in X-ray binaries may be somewhat different. Indeed, although the flaring/jet emission observed is generally similar to that observed in X-ray binaries, there are a number of differences, notably the persistence of radio emission in the plateau phase (peak of outburst) of dwarf novae outbursts. Further, novalikes, which have persistently high accretion rates have also been shown to have jet emission e.g. \cite{cope20}.

Emission corresponding to jets has also been detected from several other cataclysmic variables, see \cite{cope20} for a review. However, one system, VW Hyi, has not shown evidence for jets, even though it is very nearby ($\sim$54 pc \cite{pala19}) and the reason for this is unclear. It may be that the jet emission, which is transient, was simply missed. Alternatively, it may be due to the mass of the white dwarf in this system which is significantly less massive than the white dwarfs in the other systems exhibiting jet emission, where the white dwarf in VW Hyi is 0.67 M$_\odot$ and those in the other systems have masses $>$0.8  M$_\odot$
with the white dwarf in U Gem measuring 1.2  M$_\odot$ \cite{ritt11}. This lower mass and thus larger radius (see Section~\ref{sec:chandra_mass}) will influence the accretion rate, where a lower accretion rate implies less accretion power, which could lead to weaker jets or even no jets.

Magnetic cataclysmic variables have also been seen to show radio emission reminiscent of jets, but it is not yet clear if this radio emission is coming from a jet or from somewhere else within the system (secondary star, interacting matter between the two stars, etc) \cite{cope20}.

\subsection{Binary components and the diversity in accreting white dwarfs}

The diversity in the white dwarf masses and magnetic fields, as well as the diversity in the companion star mass and nature (main sequence or evolved star or compact object) along with the evolutionary stage of the binary, gives rise to a very wide range of accreting white dwarf systems. This diversity is covered in the following sections.

\subsection{Cataclysmic Variables}
\label{sec:cvs}

\begin{figure}[h]
\centering
\includegraphics[width=12cm]{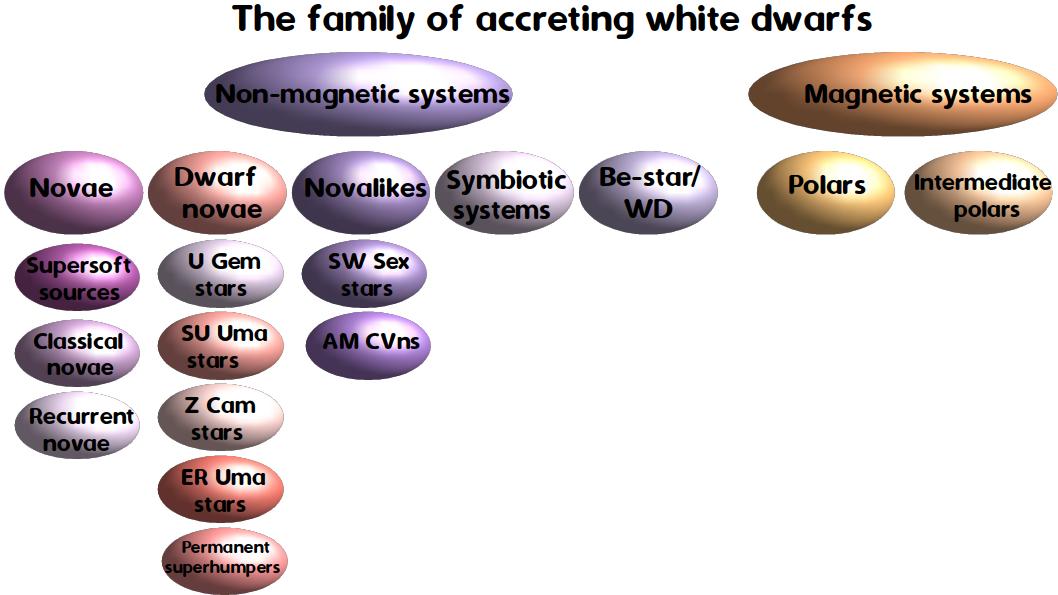}
\caption{Different classes of accreting white dwarfs and their sub-classes.}
\label{fig:CVblock_diag}
\end{figure}

The term \hl{Cataclysmic Variable} (CV) describes a wide variety of semi-detached close binary systems, where the primary is a white dwarf, accreting matter from a companion (secondary) star via Roche Lobe overflow (see Section~\ref{sec:acc}  and Chapter~\ref{ch:accretion}).  A block diagram showing the different classes and sub-classes of accreting white dwarfs is given in Figure~\ref{fig:CVblock_diag}. CVs are considered non-magnetic when the field is not strong enough to alter the accretion flow, often taken to mean that the \hl{accretion disc} extends all the way down to the surface of the white dwarf.

\begin{figure}[h]
\centering
\includegraphics[width=15cm]{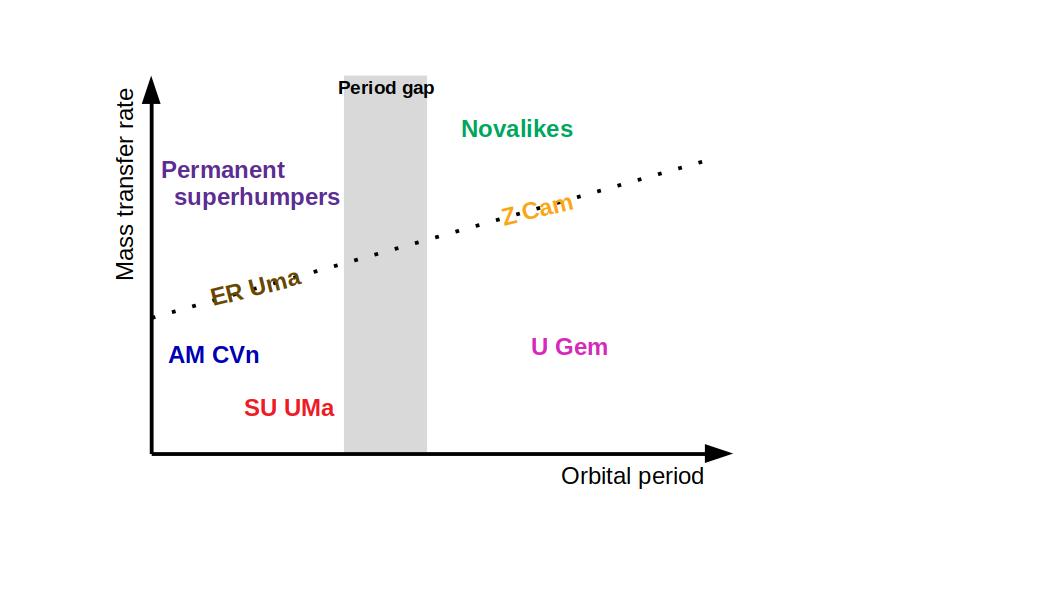}
\caption{Different classes of dwarf novae shown with respect to their approximate  mass transfer rate as a function of orbital period.  The period gap is indicated as a grey vertical bar and the critical mass transfer rate (see Section~\ref{sec:dn_nl}) is indicated as a dotted line.}
\label{fig:cvdiag}
\end{figure}

The sub-classes of CVs, were originally defined from the shape of the optical photometric lightcurves. Nowadays, spectroscopic and polarimetric \hl{observations} are used to classify the various CVs into sub-classes. Many of the sub-classes can be categorised simply by the mass transfer rate
and the type (if any) of \hl{outbursts} that they exhibit. By plotting the
mass transfer rate against the orbital period, CVs can
be shown to fall roughly into different categories \cite{osak96} and shown in
Fig.~\ref{fig:cvdiag}.  The main features
of each category are further described in the following subsections.

\subsubsection{Classical novae}
\label{sec:CN_sss}

\hl{Classical novae} have been known for hundreds (or maybe thousands) of years as they can be so bright, they were initially mistaken for new (nova) stars appearing in the Universe, which explains the origin of their name. Classical novae are now known to be luminous transients, brightening in the optical by up to 9 magnitudes in a few days. This occurs when degenerate shells of accreted material burn explosively (thermonuclear runaway) on the surface of white dwarfs. The speed of the decline phase gives the nova class. These are usually expressed as either t$_2$ or t$_3$, indicating the time to decay by either two or three magnitudes from the peak magnitude. The classification is then given by $very\ fast <$10 days, $fast$ 11-25 days, all the way to $very\ slow$ 151-250 days. Shorter decay times usually indicate brighter novae.

In general, nova progenitors are novalike binaries,  exhibiting high rates of mass transfer to their white dwarfs before and after an eruption. At the outset of a Classical nova, there is an optically thick, high velocity wind and the mass loss decreases with time. As the photosphere recedes, the degree of ionisation increases giving rise to stronger and stronger, higher emission lines, coming from successively deeper shells with lower and lower velocities. Thus the emission lines narrow. Absorption lines come from two distinct shells that collide. The ejecta are enriched in carbon, nitrogen, oxygen and neon \cite{cass04}. The typical mass of the ejected material, the nova shell, is estimated at 10$^{-5} - 10^{-4}$ M$_\odot$ \cite{yaro05}. The outburst is so powerful that the nova shell is often observed for a long time after the eruption.

The nature of the white dwarf, either carbon-oxygen or oxygen-neon, plays a role in the type of ejecta. Analysing the ejecta can give insight into the nucleosynthesis occurring during the outburst.   As the radius of pseudo-photosphere shrinks, the peak of the emission shifts from the optical to soft
X-rays.  For some novae the photosphere does not recede to the regions within the outflow that are hot enough to produce X-rays and they move directly to what is called a supersoft source phase (described in the following). Alternatively, they can move to this phase after an X-ray bright phase.

Some classical novae are recurrent, outbursting every few decades or so. Some novae can also stem from symbiotic systems. \\

{\bf Supersoft sources} \\

Following the classical nova explosion, a fraction of the envelope is ejected and the expanding shell becomes optically thin to X-rays, giving rise to a bright source of supersoft X-rays in some sources. This is powered by residual hydrogen burning continuously on the surface of the white dwarf.  As the envelope mass is depleted, the photospheric radius decreases at constant bolometric luminosity (close to the Eddington value, the luminosity at which the force due to radiation pressure equals the gravitational force from the mass of the star, supposing spherical accretion, and thus stopping a higher accretion rate and thus higher luminosity, $\sim$ 10$^{38}$ erg s$^{-1}$ for white dwarfs) with an increasing effective temperature. This leads to a shift of the spectral energy distribution from the optical to the soft X-rays. This is known as a supersoft source (SSS) \cite{grei96} and it has a hot white dwarf atmosphere spectrum. They were first discovered with the {\it Einstein} observatory, in the Large Magellanic cloud \cite{long81}.

The duration of this supersoft source phase is related to the nuclear burning timescale of the remaining hydrogen rich envelope and depends, among other factors, on the white dwarf mass \cite{sala05}. The typical duration for this phase is a few months \cite{piet05}.  Low resolution X-ray spectra are well fitted with blackbodies with temperatures of the order of 10$^5$ K (emission $\lesssim$ 1 keV), but higher resolution X-ray spectra show broad absorption lines such as carbon, oxygen and nitrogen, see Chapter~\ref{ch:white_dwarfs}.

\subsubsection{Dwarf novae and novalikes}
\label{sec:dn_nl}

The term \hl{dwarf nova} is an all encompassing term for cataclysmic variables that show (semi-)regular outbursts. These outbursts are fainter than those observed from classical and recurrent novae. There are many types of dwarf novae, which are described below. The novalikes set themselves apart from the other dwarf novae as they are almost always in a hot state, similar to the novae outbursts. However, they can occasionally be seen to fade as described below. \\

{\bf U Gem stars} \\

U~Geminorum (U~Gem) stars have orbital periods longer than the top end of the period gap (see Section~\ref{sec:periods}) and have
low mass transfer rates (low ${\rm \dot{M}}$), see Fig.~\ref{fig:cvdiag}.  The secondary star is
usually a K- or early M-dwarf \cite{ritt11}.  U~Gem stars
exhibit regular outbursts, primarily detected in the optical domain, where they increase in brightness by 2-5 magnitudes, for example the
outbursts of AL~Com, V544~Her, V660~Her, V516~Cyg and DX~And \cite{spog98} and see also Fig.~\ref{fig:AAVSOlc}.  U~Gem outbursts occur every few weeks or months and
last for days-weeks, depending on the size of the system and the mass
transfer rates, e.g. U~Gem \citep{sion97} and SS~Cyg \cite{cann98}.  The majority of outbursts observed from U~Gem stars are outside-in (see Section~\ref{sec:acc}), but some inside-out outbursts have also been observed, e.g. \cite{webb99}.  The quiescent
lightcurves of eclipsing systems, show a brightening prior to eclipse,
as the bright spot (the region where material from the secondary
impacts the accretion disc) rotates into view.  The spectra of
U~Gem stars taken in quiescence, show Balmer and helium emission lines
which are common for dwarf novae in general.
The emission lines can show a double-peaked profile, consistent with
an accretion disc. \\

{\bf SU~UMa stars} \\

\begin{figure}[h]
\includegraphics[width=12cm]{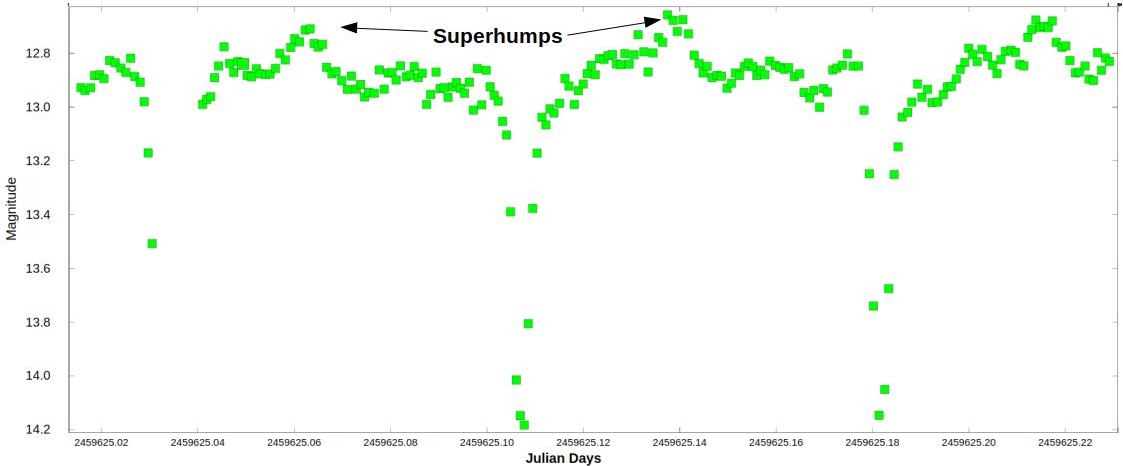}
\caption{The V-band lightcurve of the cataclysmic variable Z Cha, taken from the AAVSO database. The data were taken on 14th February 2022. Three eclipses of V$\sim$1.4 depth are visible in the 5 hours of data. Shortly after each eclipse, a $\sim$0.2 magnitude brightening is visible. These are known as superhumps.}
\label{fig:AAVSOlcZCha}
\end{figure}

SU~Ursae Majoris (SU~UMa) stars have periods below the period gap (see Section~\ref{sec:periods}), or $\lesssim$2 hours and
have low $\dot{M}$ \cite{osak96}, see Fig.~\ref{fig:cvdiag}.  The secondary is usually a
late-ish M-dwarf \cite{ritt11}.  SU~UMa systems exhibit
semi-regular outbursts, but also have occasional super-outbursts \cite{wagn98}.  In a super-outburst, the system brightens
slightly more than it would in an ordinary outburst (up to about a
magnitude brighter), but the outburst lasts about 5 times longer than
an ordinary outburst.  SU~UMa systems are theoretically predicted to
show only `outside-in' outbursts \cite{smak84}.  Evidence of the bright spot in the quiescent photometric
lightcurve is observed, e.g. OY~Car \cite{wood89b} and the spectra
show emission lines typical of dwarf novae.
SU~UMa stars can also be recognised by `superhumps' seen in their
photometric lightcurves, during super-outburst, e.g. in VY~Aquarii \cite{patt93} and Fig.~\ref{fig:AAVSOlcZCha}. Due to tidal instabilities, where the gas in the disc
experiences the gravitational force of the secondary, a
brightening (hump) in the lightcurve is observed.  This occurs in
systems where the mass ratio is $0.03 < q < 0.33$, where $q$ is the
mass of the secondary ($M_2$) divided by the mass of the primary
($M_1$).  The period of the hump is greater
than the orbital period by a few percent.  `Negative' superhumps have
also been observed, where their period is less than the orbital period
of the system e.g. \cite{pavl21}. \\

{\bf Z~Cam stars} \\

Z Camelopardalis (Z~Cam) stars have orbital periods $\gtrsim$3 hours (above the period gap) and have a
higher mass transfer rate than either the SU~UMa or the U~Gem stars, e.g. TZ~Per \cite{ring95}.  The mass transfer rate is
approximately the critical mass transfer rate, above which the system
is stable and no outbursts occur (as with the novalikes).  Below the critical mass transfer rate
the system is unstable and the system exhibits outbursts, as in the
U~Gem stars.  Therefore, they exhibit
outbursts and are also included in the dwarf nova group, but these in turn can exhibit standstills, e.g. \cite{hone98}.  A standstill can occur on the decline from outburst.
As the brightness of the disc fades, it can sometimes be seen to stall
and remain at the same brightness ($\sim$0.7 magnitudes fainter than
the peak of outburst) for weeks to years, e.g. in AT~Cnc \cite{noga99}.  This is because the mass transfer rate has increased above a
critical rate and the system has entered a quasi-stable
state.  Until the mass transfer rate drops, the system will remain in
standstill.  Photometry of the standstills reveal small fluctuations
in brightness and some flickering e.g. \cite{noga99}.
During standstill, the spectra of the Z~Cam stars show the Balmer jump
and the Balmer series in absorption e.g. \cite{noga99}.
He~II (4686\AA) can also sometimes be seen in emission. \\

{\bf ER UMa stars} \\

ER Ursae Majoris (ER~UMa) stars share many of the same characteristics as the SU~UMa
stars, as they have an orbital
period below the period gap ($\lesssim$2 hours).  They are indeed often thought to simply
be a sub-class of the SU~UMa stars, e.g. \cite{gao99}.  Like the
Z~Cam stars, they have a mass transfer
rate that lies near to the critical mass transfer rate, see
Fig.~\ref{fig:cvdiag}, and so they also show standstills in some of
their outbursts \cite{kato95}.  Superhumps, however, are not
just observed in the photometric lightcurve during superoutburst, but have also
been detected during `normal' outbursts, e.g. ER~UMa \cite{gao99}. \\

{\bf Permanent superhumpers} \\

These systems have short orbital periods, but have a mass transfer
rate above the critical mass transfer rate e.g. \cite{rett00} and see
Fig.~\ref{fig:cvdiag}. They therefore do not exhibit outbursts.  There is evidence for a
permanent superhump in their photometric lightcurves. \\

{\bf Non-magnetic novalikes} \\

There are many sub-classes of novalikes, but in general, they have orbital periods longer
than $\sim$3 hours and have a higher mass transfer rate than even the
Z~Cam stars.  They are therefore in a
stable state and have never been observed to go into outburst.  Little
or no evidence of a bright spot is seen in the photometric or
spectroscopic lightcurves.  The optical/UV spectra show a larger number of
high-excitation emission lines than the other cataclysmic variable, e.g. UU Aquarii \cite{bapt00}.  At these high accretion rates, the novalikes (like dwarf novae in outburst) have an optically thick boundary layer. In this region close to the surface of the white dwarf, the gas moving at Keplerian velocities is decelerated
to the same speed as the surface velocity of the white dwarf.  Some of the gas arrives at the surface
of the compact object and spins it up through viscous shear stress \cite{stoe80}, but
the rest, is simply dissipated in
the boundary layer.  As the boundary layer is just a small thin
region, the radiation emitted from this region is high energy. Depending on the nature of the boundary layer, whether it is optically thin or optically thick, and the rotation
rate of the compact object, the radiation emitted can be from the far ultra-violet to
hard X-rays.  A  radiation-driven accretion
disc wind is also a common phenomenon among novalike CVs.

For orbital periods between three
and four hours, a class of CVs exist called SW Sex stars. These systems show single Balmer emission lines as well as a prominent
HeII 4686 \AA\ emission line. They show transient absorption at some phases, and are
often discovered as eclipsing systems. A few SW Sex stars show low circular polarisation, indicating
a relatively high magnetic field for the white dwarf. Some novalikes, including the SW Sex stars, occasionally
fade by several magnitudes, for weeks to years. During these low states, the
mass transfer from the secondary star is interrupted, and the accretion disk
becomes very faint, or vanishes altogether. Consequently, the white dwarf and
the secondary star can be observed directly. The white dwarfs in
these systems appear to be very hot, which suggests that the high accretion rates
observed during the high states could persist for centuries \cite{szko12}. \\

{\bf AM CVn binaries} \\

AM Canum Venaticorum (AM CVn) stars are a class of compact binaries in
which the white dwarf accretes hydrogen-poor and helium-rich material
from a degenerate (or semi-degenerate) companion. These objects are
characterised in the optical/UV by their strong helium spectral lines;
their lack of (or very weak) hydrogen features and by their short
orbital periods ($<$70 min), see Fig.~\ref{fig:CVperiods}.

Many AM CVns show superoutbursts aswell as regular outbursts. These superoutbursts are generally similar to those seen in the short period SU UMa stars. During the
superoutbursts, these systems also show superhumps. Thanks to their strongly variable nature, recent sensitive large sky surveys have revealed a significant population of these systems, e.g. \cite{pich21}.

These systems are expected to be persistent gravitational wave sources that will constitute a large proportion of the gravitational wave emission detected by the future observatory {\it LISA} (Laser Interferometer Space Antenna), see also Section~\ref{sec:future_survey}. They will therefore be useful calibrators for the mission and locating them and determining their periods, component masses etc will be essential for fully exploiting this mission. A larger population will also be discovered with LISA, with 2200-5200 expected in our galaxy \cite{liu21}.

\subsubsection{Other non-magnetic CVs}
\label{sec:otherCVs}

{\bf Symbiotic stars} \\

Symbiotic stars are wide, interacting binaries in which a compact object accretes from a red giant companion. This section will focus on the systems in which the compact object is a white dwarf. Due to the larger size of the companion, the binary separation is larger than in the systems discussed above, spanning a few to hundreds of astronomical units, thus the orbital periods range from years to decades.  They were first discovered in 1932, thanks to optical spectroscopy revealing the TiO bands seen in the red giant spectra, along with emission lines from the binary interaction \cite{merr32}, though at that time, their nature remained unexplained.

Matter is lost from the red giant to the white dwarf via the inner Lagrangian point, either via a wind or through Roche lobe overflow. In some symbiotic stars, the emission is generated by the
release of gravitational potential energy as matter falls onto the white dwarf. For the others, the emission arises primarily from the quasi-steady burning
of a hydrogen-rich shell on the white dwarf surface e.g. \cite{soko17}. The systems are highly variable over the orbital period in all wavelengths.  Ellipsoidal modulation in the near infra-red is seen if the red giant fills its Roche lobe. In this case, more emission is seen when the star is viewed side on, presenting a tear drop shape, than when seen from behind, when it appears spherical.  Modelling this modulation can provide the inclination of the system, as a face on system (inclination, $i$=0$^\circ$) shows no modulation and a side on system ($i$=90$^\circ$) shows maximum modulation. Strong irradiation of the matter between the companion and the white dwarf gives rise to strong Balmer emission lines, which vary significantly as they rotate in and out of eclipse. In X-ray, they can show similar low resolution spectra to the supersoft sources or harder sources where the boundary layer is detected. Some show evidence for colliding winds between the two components.

The systems also show longer term variability, showing outbursts as in the cataclysmic variables and some have shown classical nova eruptions. In symbiotic stars, the equivalent of polars should not exist due to the much larger binary separation, but there is no obvious reason why IP-like systems should not exist.

These systems are particularly interesting to study as they are wide and therefore the timescales are much longer than in the systems discussed above, making them easier to study. These systems are also potential progenitors of \hl{type 1a supernovae}. \\

{\bf Be star-white dwarf systems} \\

In these systems, the white dwarf accretes from a B star showing emission lines due to a circumstellar disc. Accretion usually occurs as the white dwarf, which is in an elliptical orbit, passes through the disc at periastron.  Few such systems are known to date, although they are predicted to exist in large number \cite{ragu01}. These systems are particularly interesting from an evolutionary point of view as more massive stars evolve more rapidly than lower mass stars. As seen above, the maximum mass of a star that ends its life as a white dwarf is 10.6 M$_\odot$. The companion star should therefore be less massive than this mass, ruling out a large number of the B stars.

Be-white dwarf binaries often show transient (as the white dwarf passes through/ near the disc), soft X-ray spectra, similar to the supersoft sources. However, the $\gamma$-Cas systems, also thought to be Be star binaries but with very hard $\gamma$-ray spectra, may also be a part of this family, e.g. \cite{motc15}.

\subsubsection{Magnetic CVs}
\label{sec:polsandips}

{\bf Polars and Intermediate Polars} \\

These \hl{magnetic} CVs used to be classified together with the novalikes.
However,  when AM Her was shown to be linearly and circularly
polarised \cite{tapi77}, where previously only magnetised white dwarfs
had been observed to be polarised, a new sub-class of
magnetic cataclysmic variables was realised.  Since then, many other CVs with
magnetised white dwarfs have been discovered. When the magnetic field of the white dwarf is strong enough to completely disrupt the accretion disc, with fields typically of B$\gtrsim$10$^7$ G, they are known as the
AM~Her~stars or polars. Polars accrete without an intervening accretion disk. The matter lost from the inner Lagrangian point of the
secondary follows a ballistic trajectory, and is subsequently captured by the magnetic field of the white dwarf. It then follows the field lines onto the poles of the white dwarf. This material heats up through strong adiabatic shocks and cools subsequently through plasma emission in the X-ray domain and optical cyclotron radiation.  The field magnetically locks the white dwarf spin period to the orbital period, so all variations are seen on the orbital timescale.

Asynchronous polars can be found, but the difference between the orbital and spin period are generally $<$1\%. The other properties are common with those of polars. 
For white dwarf magnetic fields of 1$\lesssim$B$\lesssim$10 MG, the acretion disc is only partially disrupted, in the central regions, so material that is accreted onto the white dwarf, does so via
magnetic field lines that connect the accretion disc to the white
dwarf. These systems are known as intermediate polars (IPs) or DQ Her type systems \cite{patt94}.  The white dwarf magnetic moment in this case is insufficient to tidally lock the two stars and the white dwarf rotates at higher frequencies than the orbital frequency of the binary system \citep[e.g.][]{warn95}. The spin period of the primary is significantly shorter than the orbital period, so these systems are often identified by pulsations on the spin period.

\begin{figure}[h]
\centering
\includegraphics[width=10cm]{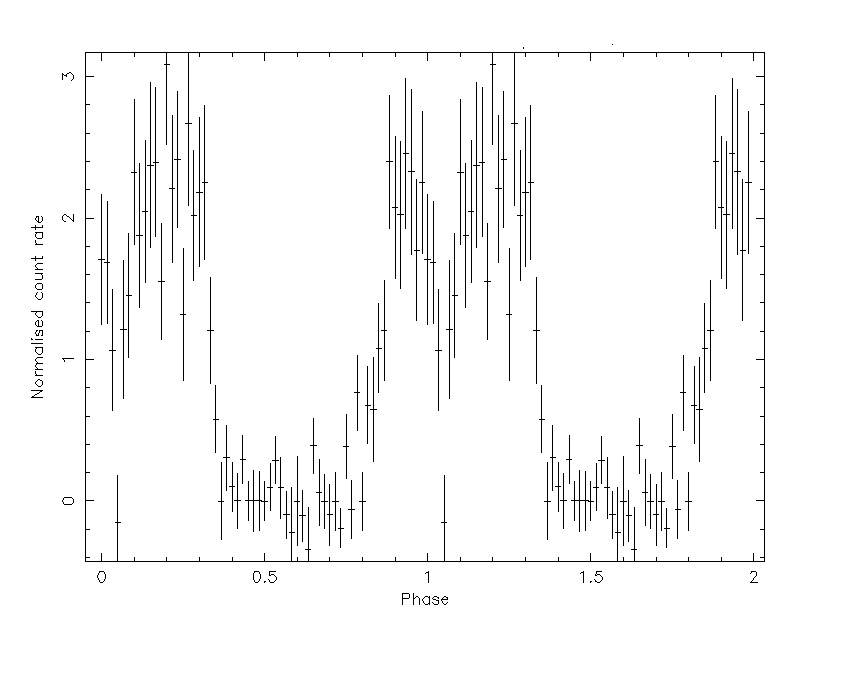}
\caption{The 0.2--12.0 keV pn lightcurve of the polar 3XMM J183333.1+225136, folded on the optical period of 7715.8 s and shown with bins of 128.6 s. Two periods are shown for clarity.  Only one pole is visible in this system, causing the bright phase between from phases $\sim$0.8-1.2. An eclipse of the white dwarf is seen at phase 0.05. Courtesy of \cite{webb18}.}
\label{fig:PolarLightCurve}
\end{figure}

The secondary stars of the polars and
intermediate polars (IPs) are late-type stars \cite{ritt11}. The photometric lightcurves show that these magnetic CVs tend to hover
about a maximum brightness, and occasionally, after years, fall to a
lower brightness, e.g. AM~Her itself \cite{wats80}.
Some of the IPs, however, have not yet been observed to show a low
state.  There is also flickering, some flaring and quasi-periodical
oscillations evident in the lightcurve, e.g. AM~Her \cite{beue91}.  The optical/UV spectrum shows many emission lines \cite{patt94}. In the X-ray, the lightcurve shows either one or two bright pulses per orbital period, depending on whether one or both poles are visible, when the heated polar column becomes visible due to the viewing angle, see for example Fig.~\ref{fig:PolarLightCurve} and Chapter~\ref{ch:white_dwarfs}.  A single wide sinusoidal ``pulse'' can be visible if one pole is always visible. A central, short duration, dip due to the eclipse of the white dwarf is seen if the inclination is  $\gtrsim$75$^\circ$.  The short duration is due to the small size of the white dwarf.  With regards to the X-ray spectra, these vary strongly from polar to polar, but with sufficient counts and resolution, multi-temperature shock models with complex absorbers fit the data at higher energies and some polars show very strong soft excesses. Many polars also show an iron emission line at 6.4 keV, with equivalent widths of $\sim$150 eV \cite{muka17}. One way in which IPs are often identified is through the modulation of the emission on the spin period of the white dwarf. A third period, in addition to the orbital period (P$_{orb}$) and the spin period (P$_{spin}$) is sometimes observed. This is the beat period between the two periods, such that 1/P = 1/P$_{spin}$ - 1/P$_{orb}$.

Magnetic CVs are interesting to study as the mass, radius and composition of the white dwarf can be constrained \cite{muka90}, and the accretion processes can be probed \cite{rams04}.  Further, short orbital period CVs are often faint in the optical domain, but can be bright in X-rays if they are magnetic. Using the high energy domain can then be helpful for identifying CVs with the shortest orbital periods close to the period bounce, to help understand how such objects evolve.

\subsection{Accreting white dwarfs in the broader astrophysical context}

\subsubsection{The origin and evolution of accreting white dwarfs}
\label{sec:binary_evol}

Binary systems containing accreting white dwarfs often evolve from a primordial binary systems composed of stars of $\lesssim$8 M$_\odot$ masses, see Section~\ref{sec:WD_form} and \cite{iben85}. The most massive star will generally evolve first to become a white dwarf, passing through the red giant phase, that if the two stars are close enough, will lead to a common envelope, which will permit angular momentum to be removed from the binary and allow the binary to harden (the two stars will become closer), see Chapter~\ref{ch:binary_evol}. This may allow the two stars to become semi-detached, see Section~\ref{sec:acc}, allowing them to interact as a Cataclysmic variable, see Section~\ref{sec:cvs}. Depending on the relative masses of the component stars, the binary separation, the white dwarf magnetic field strength, the local environment, as well as other characteristics, the different types of CVs, see Section~\ref{sec:cvs}, will be formed. Later, however, the companion star can evolve through the red giant phase, causing higher rates of accretion as te star overfills its Roche lobe, or the system to become a symbiotic, see Section~\ref{sec:otherCVs} and possibly a second common envelope phase, removing further angular momentum from the system. Eventually, the binary will become a double white dwarf system, see Section~\ref{sec:dn_nl}, and again depending on the relative masses of the component stars, the binary separation, the white dwarf magnetic field strength and the local environment, accretion may continue, see Section~\ref{sec:type1a}.

Alternatively they can form binaries through tidal capture of a star in a dense stellar environment, such as globular clusters (see Section~\ref{sec:GCs}) or galactic centres. Binaries can subsequently encounter neighbouring stars. If the binding energy of the binary is larger than the kinetic energy of the star, angular momentum will be removed from the binary and it will harden.  Binaries can undergo a number of such encounters, which will have a strong impact on their evolution. The third body leaves the encounter with the extra energy extracted from the binary e.g. \citep{hut92}. This significantly modifies the evolutionary path of the binary, see for example \cite{ivan06}, creating different population sizes of the same types of white dwarf binaries created through primordial binary evolution, described above.

\subsubsection{Observed orbital period distribution}
\label{sec:periods}

\begin{figure}[h]
\centering
\includegraphics[width=12cm]{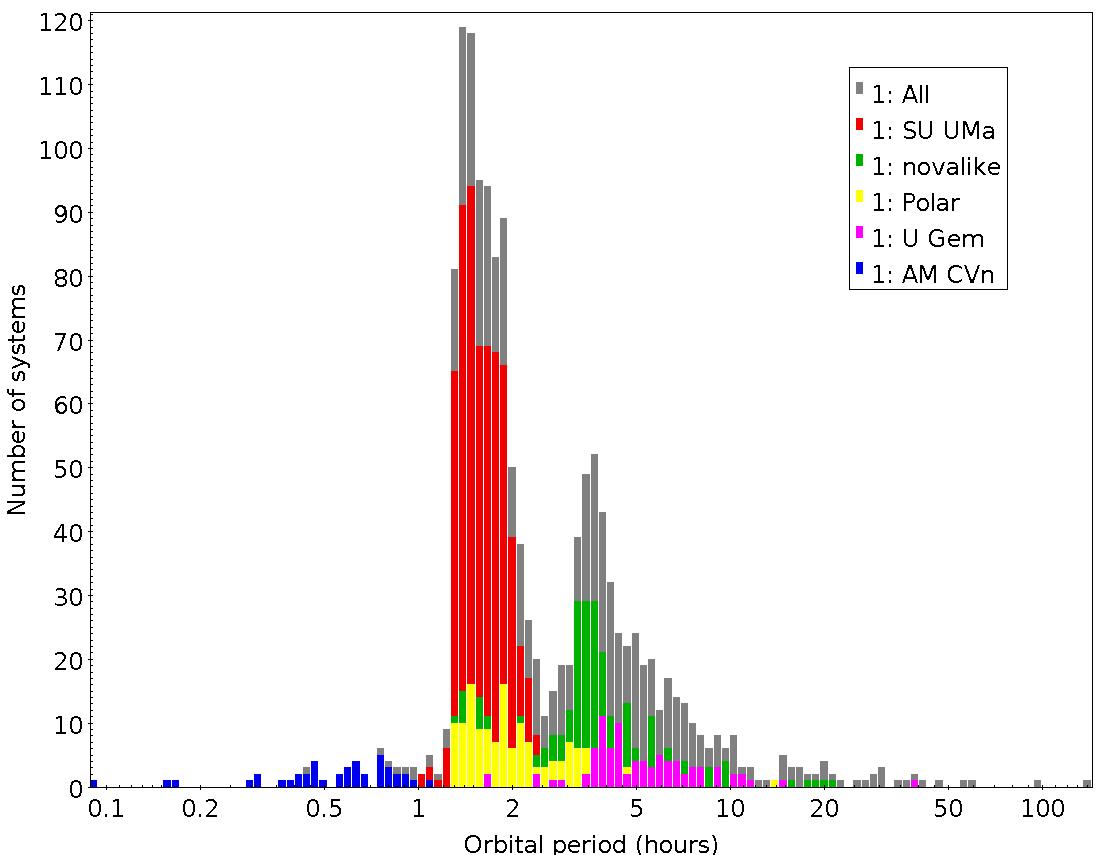}
\caption{Distribution of the orbital periods of white dwarf binaries taken from \cite{ritt11}. The period gap is clearly visible between $\sim$2.3 and $\sim$2.8 hours, as is the period spike at around 1.2 hours. Different types of systems are indicated, when the nature is known. The polar category also includes intermediate polars.}
\label{fig:CVperiods}
\end{figure}

White dwarf binaries have a very wide range of periods, dictated by the mass of the two components and the separation that allows them to be semi-detached. In reality, the Roche lobe filling star’s mean density, $\rho_\odot$, is almost wholly responsible for determining the orbital period (in hours) $P_h$, where  $P_h = \kappa/\rho_\odot^{0.5}$ (where $\kappa \simeq$ 8.85 is only a weak function of the mass ratio) e.g. \cite{kolb99}.  As can be seen in the Fig.~\ref{fig:CVperiods}, the orbital periods can be as long as 5 days and as short as 5.4 minutes. For the shortest periods, these are the AM CVn systems, see Fig.~\ref{fig:CVperiods} (the blue bars). The secondary star in AM CVns is either a helium rich white dwarf or a helium rich semi-degenerate star and therefore has a small radius, allowing such short orbital periods. However, for CVs with main sequence secondary stars, the periods are longer. Using the above equation, a minimum period of $\sim$70 minutes is calculated. In the distribution of periods shown in Fig.~\ref{fig:CVperiods}, the lower bound to the bulk of the CVs (which are not AM CVn type systems) is $\sim$72 minutes, commensurate with this value. \\

{\bf The period spike} \\

The donor star density on the main sequence increases as it tends towards lower masses, as it loses matter to the white dwarf. Nearing the hydrogen-burning minimum mass, the increasing electron degeneracy brings around structural changes so that continued mass loss reduces the density. Therefore as the donor star transitions from a main-sequence star to a brown dwarf, the mean orbital period derivative changes from negative to positive (passing through zero at period minimum, i.e. \.P$_{orb} = 0$), so the period begins to increase instead of steadily decreasing e.g. \cite{kolb99}. This means that there should be a large number of CVs around the period minimum, known as the {\em period spike}. This is indeed seen in Fig.~\ref{fig:CVperiods}. However, X-ray luminosity functions imply a larger, as yet unseen population of these faint, short period CVs, e.g. \cite{pret12}. The shortest period CVs correspond to CVs with the lowest mass white dwarfs \cite{ritt11}.\\
 
{\bf The period gap} \\

There is
a dearth of CVs, including IPs, with periods between
approximately 2.3-2.8 hours, and this is called the period gap.  It is thought to arise due to disrupted magnetic braking.  Magnetic braking (see Chapter~\ref{ch:binary_evol}) is responsible for driving the evolution of CVs with periods greater than approximately 3 hours. As the rotation of the secondary is slowed, angular momentum is lost from the binary and the companion expands slightly to maintain thermal equilibrium. At masses of 0.2-0.3M$_\odot$, stars lose
their radiative cores and become fully convective \cite{gros74}.  The magnetic field in solar-type, low-mass stars is
expected to be anchored at the interface between the radiative core and the convective envelope (the tachocline, \cite{char97,macg97}). Since fully convective stars do not
possess such an interface, the magnetic field and thus the magnetic braking, are assumed to vanish at this point \cite{knig11}. It is then the gravitational radiation that drives the evolution, which at these periods is less efficient, so the angular momentum lost from the system and the mass transfer rate drop.  The secondary will shrink back to its original size and will thus be inside the Roche lobe, so the binary becomes non-detached.  Through the emission of gravitational radiation removing energy from the binary system, the
separation between the two stars could decrease \cite{rapp83} and the secondary could eventually come into contact again
with its Roche lobe.  Mass transfer could then recommence \cite{rapp83}.  This would happen at an orbital period of
about 2.3 hours \cite{king88} and hence CVs are generally observed above and below the period
gap and not within it.

This would then imply that the companion stars just above and below the gap should have identical masses, but different radii.  Indeed \cite{patt05} showed that there is a clear discontinuity in donor radii at $\simeq$ 0.2M$_\odot$ at the interface between the longer-period systems from the shorter-period systems. Indeed, donors in systems just below the period gap have radii consistent with regular main sequence stars of this mass, while the donors just above the gap have radii that are inflated by $\simeq$ 30\%. This gives excellent support to the disrupted magnetic braking model.

Some magnetic systems have periods that fall within the period gap. As the white dwarf in a polar rotates synchronously with the orbital period, so the secondary star field lines that are normally
open in other binary systems, become
connected with the field lines of the magnetic white dwarf.
This material, that would normally be forced to co-rotate with the
secondary out to large radii and hence cause magnetic braking, becomes trapped in what is known as a
`dead zone' \cite{wick94}.  In such a way, magnetic
braking cannot be an efficient loss of angular momentum in polars.  The orbital evolution of polars should
then be driven mainly by angular momentum loss due to
gravitational radiation \cite{wick94}. This scenario
explains the higher proportion of polars in the period gap when
compared with other CVs, as there is no sudden cut-off in magnetic
braking.

\subsubsection{Exceeding the Chandrasekhar mass}
\label{sec:type1a}

As seen in the Section~\ref{sec:chandra_mass}, the maximum mass for a white dwarf is the Chandrasekhar mass. If the white dwarf accretes matter such that the mass exceeds this limit, the electron degeneracy is not sufficient to support the gravitational force due to this mass. Exceeding this mass implies that the core will shrink further and the density will increase. At high densities, of the order of 10$^7$ g cm$^{-3}$, the highest energy electrons have energies greater than the difference between the masses of the neutrons (m$_n$) and the protons (m$_p$), m$_n$c$^2$ - m$_p$c$^2$ = 1.294 MeV. At this time it becomes energetically favourable for the protons to capture electrons via the weak-interaction i.e. for inverse $\beta$-decay to occur (Eq.~\ref{beta_decay})

\begin{equation}
e^- + p \rightarrow\ n + \nu_e
\label{beta_decay}
\end{equation}
where $p$ is the proton, $n$ is the neutron and $\nu_e$ is the electron-neutrino.  The inverse $\beta$-decay generates more neutrons than electrons and protons due to the fact that the electrons still in the star are degenerate and have filled all the available states, stopping new electrons from being created via neutron $\beta$-decay. Thus a neutron star is created.  This is known as accretion induced collapse. At birth, the nucleons have free fall energies, $\sim$100 MeV per nucleon ($\sim 10^{12}$ K). The neutron star cools quickly through neutrino emission, so that after just a few seconds it is $\sim 10^{11}$ K \cite{mill13}. The core continues to cool in such a way. This is known as the {\it Urca} (name originating from the place at which the process was elaborated) process (or Durca for direct Urca):
\begin{equation}
n \rightarrow\ p + e^- + \bar{\nu_e}
\label{eq:urca}
\end{equation}
so the neutron star will have a temperature of the order of 10$^9$ K within years. Over approximately a century, the surface temperature stabilises at about 100 times the temperature of the core.

Before this can occur, carbon burning can take place. This process creates a large instability, which at some point can turn into a runaway. The burning then proceeds very rapidly through the star and can disrupt it completely within seconds. Depending on how the burning proceeds,  a luminous supernova, known as type Ia supernova, can be produced, e.g. \cite{goob11}.  This process can be modelled and gives rise to a standard luminosity for the supernova. Thanks to their high energy ($\sim$10$^{46}$ J), they can be detected out to cosmological distances. As the luminosity (L) is standard for these events, the distance (d) to the supernova can be determined from the measured flux, and knowing that the flux is L/(4$\pi$d$^2$). The distances can then be used to understand the local expansion rate of the Universe as well as its structure. It is therefore important to understand how they occur, whether it is simply through the accretion onto the white dwarf or through the coalescence of two white dwarfs e.g. \cite{iben85}, how the burning proceeds \cite{goob11}, etc.

Type 1a supernovae are detected at all wavelengths and the emission stems from the radioactive decay of newly synthesised material, as well as Compton scattering. 

Despite this theory, there has been much debate in the literature as to whether a white dwarf can indeed accrete so much mass as to reach the Chandrasekhar mass (or close to this value). This is because if mass transfer is too slow, novae occur, which appear to remove much of the accreted mass. If the mass transfer is higher, hydrogen burns stably, but only a small part avoids expansion and mass loss  e.g. \cite{vank10} and references therein. However, the average white dwarf mass in accreting white dwarf binaries listed in \cite{ritt11} is 0.82 M$_\odot$ with a dispersion of 0.23, with the lowest mass given of 0.16 M$_\odot$ and the highest 1.95$\pm$0.3 M$_\odot$ (CH UMa) which is massive even for a neutron star (M$_{NS} > 1.4$ M$_\odot$) and therefore maybe misidentified. The second highest is 1.55$\pm$0.24 M$_\odot$, commensurate with the maximum mass of a white dwarf, the Chandrasekhar mass (see Section~\ref{sec:chandra_mass}). All the other white dwarfs have masses between 0.4$\pm$0.1 - 1.3$\pm$0.3 M$_\odot$, with a distribution close to a Gaussian distribution.  These masses are significantly higher than those of single white dwarfs, see Section~\ref{sec:chars_WD}.

The high masses observed do not appear to be seen in CV progenitors and they can not be explained through selection effects, where more massive accreting white dwarfs are brighter and thus easier to detect. This suggests that there is indeed a mass increase in CVs. However, this is difficult to explain, as our current understanding of CV evolution and thermal timescale mass transfer suggests that such a mass increase is difficult to produce \cite{wijn15}. However, if additional angular momentum can be lost through the mass transfer process, lower mass CVs may merge, leaving preferentially CVs with higher masses \cite{pala20}. Such a model may also explain the low space density of CVs observed (compared to current theory), as well as the number of CVs observed above and below the period gap, which again is at odds with the theory and the number of period bouncers observed, also compared to theoretical predictions.

Some studies suggest that some type Ia supernovae originate from the merging of two carbon-oxygen white dwarfs e.g. \cite{dong16}. Further, mergers of white dwarfs with helium sub-giants may also produce type 1a supernovae \cite{dong16}. The future low frequency gravitational wave mission {\it LISA} will make huge advances in our understanding of the evolution of double white dwarf systems and the Galactic merger rate, how the AM~CVn systems form and evolve, the effect of tidal coupling in white dwarfs and its role in stabilising mass transfer, as well as the discovery of many new eclipsing white dwarfs \cite{mars11}, see also Section~\ref{sec:future_survey}.

\subsubsection{White dwarfs in globular clusters}
\label{sec:GCs}

Globular clusters are dense spherical systems of $\sim$10$^4$-10$^6$ old stars \citep[e.g.][]{heno61}.  Their old age implies that they should also contain many compact objects. Through dynamical evolution, globular cluster cores should become increasingly more concentrated \citep{heno61} and eventually collapse.  This core collapse should occur within  $\sim$10$^8$-10$^{10}$ years unless further energy is supplied to the central regions of the cluster.  Therefore, many Galactic globular clusters should have already undergone core collapse.  However, observations reveal that $\sim$80\% have not yet done so, see e.g. \cite{harr96}.

Globular clusters should be home to both primordial binary systems and binaries formed due to encounters in the dense environment.  The interaction of the binary with cluster stars (see Section~\ref{sec:binary_evol}) can inject extra energy into the cluster and therefore be important in helping to delay globular cluster core collapse.

The formation mechanisms of CVs in globular clusters is currently unclear. In the field, collisions are rare and therefore CVs usually evolve from their primordial binaries (see Chapter~\ref{ch:binary_evol}). However, in a globular cluster, primordial binaries may have a wide separation and as they interact with a cluster star and/or another binary, the primordial binary can break up, especially if it is located in a dense cluster region, such as the core. Alternatively, if initial binary is harder, successive encounters can cause it to harden further. Instead the binary could undergo an exchange encounter \cite{davi97}, where the encountered star exchanges into the binary if it is more massive than one of the binary stars. A final alternative is that a single white dwarf can encounter a cluster star. This encounter can raise tides on the cluster star and transfer energy and angular momentum into the envelope. If sufficient energy is transferred, the two objects will either merge or form a binary system, which may become a CV \cite{fabi75}.

Due to mass segregation, binaries sink to the centre of the cluster \cite{davi97}. Indeed the distribution of the binaries depends essentially on the central density of the cluster. A higher core density will result in a larger number of encounters, increasing the number of CVs formed through encounters, whilst  decreasing the primordial CV population. Primordial CVs are therefore expected to be found outside the core and CVs formed through encounters should be found inside the core. \cite{bell21} show that there is a bright and a faint population of CVs associated with core collapse globular clusters, and they are correlated with position in the cluster. This may provide some support to this idea.

One of the most noticeable differences between globular cluster CVs and field CVs is the lack of outbursts observed in the cluster CVs. Many types of field CVs (U Gem, SU UMa, Z Cam etc) show outbursts every few weeks to months, see above. However, the evidence for outbursts from globular cluster CVs is poor, with few outbursts detected. Amongst these are two outbursts of the CV V2 in the globular cluster 47 Tuc \cite{pare94,shar96}. Through imaging of the globular cluster M 5, \cite{shar87} detected a probable outburst and decline to quiescence of the CV V101. \cite{nayl89} confirmed the CV identification through optical observations in outburst and quiescence.  \cite{ciar90} searched for nova outbursts, as indicators for the presence of CVs in 54 of M 31’s globular clusters. With an effective survey time of about two years, no cluster exhibited evidence for a nova explosion.

It is unclear why there should be so few outbursts from globular cluster CVs, as there is no reason physical reason for why CV activity should be suppressed in these systems. \cite{shar96} suggested that tidal capture may lead to runaway mass transfer in almost all cases, which could explain the lack of CV oubursts observed. Alternatively, this may result in low accretion rates, so the time between outbursts could be long. Following the detection of two CVs in an open cluster, where one of which underwent an outburst, \cite{kalu97} noted that it would be difficult to identify a CV undergoing a similar type of outburst in a globular cluster if either photometric variability or searching for strong Balmer emission objects was used to find CVs in globular cluster centres. This could help to explain why few CVs in outburst have been detected using optical observations.
It has been suggested that the globular cluster CVs detected using X-rays may
be dominated by magnetic CVs, e.g. the 5 magnetic CVs discussed in \cite{grin99} that were detected initially by Rosat. Magnetic CVs have high L$_x$ /L$_{opt}$ ratios meaning that they are preferentially detected when observing in X-rays, and they show few outbursts because their accretion discs are disrupted by the magnetic fields \cite{patt94}. Indeed, the CVs observed with {\em XMM-Newton} for which there are X-ray spectra all have fits that are consistent with the high temperature bremsstrahlung models, typical of magnetic CVs e.g. \cite{webb06}. This could help explain why the globular cluster CVs detected through X-ray observations have not been seen to outburst.

Globular clusters may also give us insight into the evolution of cataclysmic variables, as  they may give rise to many accretion induced collapse systems \cite{ivan08}. Indeed, the slow pulsar 3XMM J004301.4+413017 in the M 31 globular cluster B091D is a good example of such a system \cite{zolo17}. Studying globular clusters may therefore help in understanding type 1a supernovae.

\section{Discovering accreting white dwarfs}

Accreting white dwarfs are often found through their strong variability e.g. \cite{pala20} in almost all electromagnetic wavebands, either when they go into outburst or through their variability over their orbital period or sometimes due to the modulation on the spin period. Due to the accretion onto the white dwarf, it can become very hot and radiate at high energies, notably in the X-ray. Extreme events such as novae can also be detected in the $\gamma$-ray domain \cite{abdo10}. Therefore, searching for high energy (UV/X-ray/$\gamma$-ray) sources can be a good way to reveal these objects, as few objects emit at such high energies. Dedicated surveys can also reveal new accreting white dwarfs. The rest of this section discusses some of the surveys that have been very successful in identifying new accreting white dwarfs and how these systems have helped to understand the nature of these objects and their effect on their environment.

\subsection{Accreting white dwarfs found in optical surveys}

\subsubsection{Accreting white dwarfs found in the SDSS survey}

The Sloan Digital Sky Survey (SDSS) \cite{york00}, carried out with the 2.5-meter telescope at Apache Point Observatory, New Mexico, continues to survey much of the northern hemisphere and has found almost 300 CVs in its Legacy survey (2000-2008) \cite{szko11}. This survey made a map in five bands u, g, r, i and z of more than 8400 square degrees of the North Galactic Cap and three stripes in the South Galactic Cap as well as more than 1.3 million medium resolution ($\sim$ 3 \AA) spectra in the band 3800-9200\AA. Small numbers of additional CVs continue to be discovered through continuing observations. The CVs were primarily identified through spectroscopic coverage of selected objects from the photometric survey and the survey  has provided an unprecedented number of CV spectra. A wide variety of systems were therefore discovered including polars, intermediate polars, novalikes, dwarf novae and objects with pulsating white dwarfs \cite{szko11}. Thanks to the spectroscopic observations, periods of many of these sources have also been determined. Given the depth of the survey and the fact that optical luminosity functions become independent of X-ray luminosity functions at (2-3) $\times$ 10$^{30}$ erg s$^{-1}$, all of the non-magnetic accreting white dwarfs should have been identified in the SDSS footprint out to 400-600 pc \cite{revn14}.

The SDSS has also been responsible for finding many of the 18 known pulsating white dwarfs in cataclysmic variables. The pulsations have been found in the photometric lightcurves. Alternatively, the newly discovered CVs were shown later to pulsate \cite{szko21b}.  Through accretion, the white dwarfs heat up and then cool again after the outburst. This allows them to transit the instability region, between 11000 and 12500 K for a white dwarf of 0.6 M$_\odot$ e.g. \cite{szko21b}, when global pulsations associated with their normal modes occur. Through asteroseismology, it is then possible to gain an insight into the interiors of these stars

\subsubsection{Accreting white dwarfs found in the Gaia survey}

The Gaia mission \cite{gaia16} aimed to detect more than a billion stars with excellent positional accuracy. Using two identical, three-mirror anastigmatic telescopes, with apertures of 1.45 m $\times$ 0.50 m, around 1.7 billion objects have been detected since 2013, with the majority of them in our own galaxy \cite{gaia18}. Amongst these, there is a significant population of CVs, including new CVs, notably discovered through their variability combined with colour and luminosity information. This has revealed 3253 CV candidates, where most are below the period gap \cite{schw21}. Gaia has also been extremely important for determining the distance to many of these objects which enables the luminosity to be determined, aiding the search for CVs and has allowed the space density of the various classes of cataclysmic variables to be determined and compared to theory, to hone our understanding of the evolution of these objects e.g. \cite{schw18}.

\cite{abri20} have been able to identify different types of CV by studying their position on Hertzsprung-Russell diagrams (luminosity versus temperature). Amongst the results, a number of period bouncers and 9 detached CVs that are currently in the period gap and have ceased to accrete (see Section~\ref{sec:periods}) have been discovered.

\subsubsection{Accreting white dwarfs found in other optical surveys}

\begin{figure}[h]
\includegraphics[width=11.5cm]{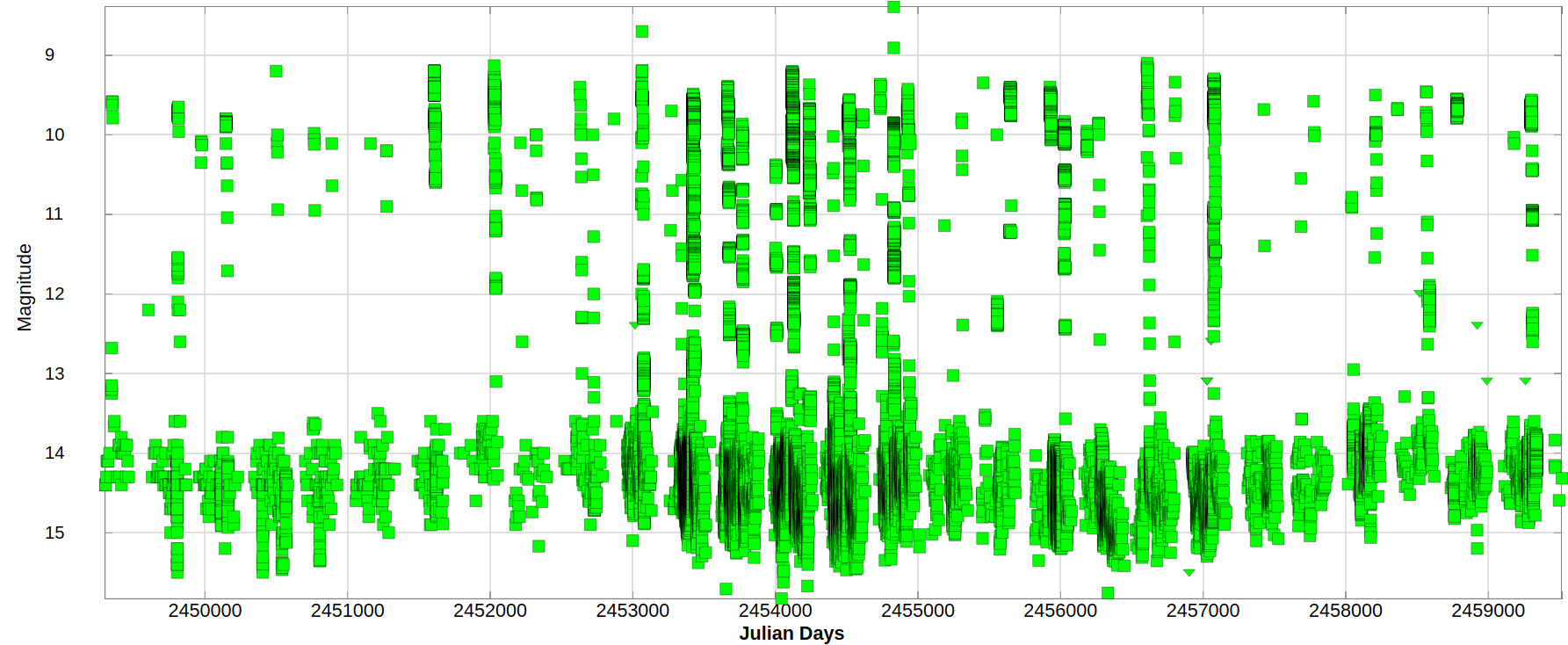}
\caption{The V-band lightcurve of the cataclysmic variable U Gem, taken from the AAVSO database. The data span May 1994 to November 2021 (27.5 years) and are made up of 57409 observations of the 204252 observations for this source in the AAVSO database. The $\sim$1.5 magnitude eclipses are clearly visible, as well as the outbursts in which U Gem brightens by as much as 5 magnitudes in the V-band.}
\label{fig:AAVSOlc}
\end{figure}

The American Association of Variable Star Observers (AAVSO) is a non-profit worldwide scientific and educational organisation of amateur and professional astronomers who are interested in variable stars. It was founded in 1911 and the members follow thousands of variable stars. This network has allowed many accreting white dwarfs to be followed over the long term, see for example Fig.~\ref{fig:AAVSOlc}, as well as  signalling the start of outbursts which have subsequently been followed up by members and professional astronomers alike e.g. \cite{webb99}. The AAVSO have also allowed simultaneous multi-band photometry of an outburst to be taken, something that can be difficult to organise on professional facilities alone e.g. \cite{webb99}. Some of the major results of the AAVSO work on accreting white dwarfs have been presented in \cite{szko12}.

Several fully-automated, wide-field surveys have operated at the Palomar observatory aimed at a systematic
exploration of the optical transient sky. This started with the Palomar Transient Factory (PTF) from March 2009 \cite{rau09},  using an 8.1 square degree camera installed on the 48 inch Samuel Oschin telescope with colours and light curves for detected
transients obtained with the automated Palomar 60 inch telescope. The intermediate Palomar Transient Factory (iPTF) used an upgraded large field camera subtending 7.8 square degrees on the sky. More recently, the Zwicky Transient Facility (ZTF, \cite{smit14}) took over with a three year, wide (field of view 47$^{\circ 2}$), three band photometric (g, r, i) survey, carried out with the Palomar 48'' telescope. Searching for blue (g-r $\leq$ 0.6) and variable (magnitude change $\geq$ 2 magnitudes) within two days, sources, ZTF have so far found 93 known and 279 new CV candidates since March 2018, at distances between 108 pc and 2.096 kpc. Most of the new CVs are at the faint end of the luminosity distribution and are therefore the lowest mass-transfer systems \cite{szko21}.

Unprecedented lightcurves of a few tens of CVs were taken over the nine years of Kepler/K2 observations (2009-2018) e.g. \cite{koch10}, which was designed to search for transits by exoplanets using optical photometry taken with the 1.4 m space based telescope. Cadences as short as 1 minute provided extremely detailed lightcurves of known and new CVs \cite{moln16}. New CVs were found through orbital variability as well as by detecting outburst from systems with short outburst intervals. As of 2018, the  Transiting Exoplanet Survey Satellite (TESS) observatory \cite{rick16} was launched and carries out a photometric (600-1000 nm) survey of a rectangular field of 24 deg $\times$ 90 degrees for 27.4 days at a time at a cadence of 30/10/2 minutes \cite{pich21}. TESS has observed many CVs with high quality lightcurves. This has enabled a detailed study of the long-term lightcurves of 9 AM CVns, showing that normal outbursts are more common than previously thought, where it was believed that superoutbursts were more common in these systems \cite{pich21} and that the duration of the latter are in fact shorter than that inferred from incomplete data. They have also allowed quasi-periodic oscillations to be discovered in the light curve of TX Col, a disc-less intermediate polar, that appear to be due to episodes of enhanced accretion, forming a torus of diamagnetic blobs near the binary’s circularisation radius. The QPOs would then be interpreted as beats between the white dwarf’s spin frequency and unstable blob orbits within the white dwarf's magnetosphere. These blobs may then lead to the formation of a disc \cite{litt21}. Many other new, as well as known CVs, symbiotics and other accreting white dwarfs have now been studied in detail, with many more results expected.

The Catalina Real-Time Transient Survey \cite{djor11} discovers dwarf novae through their optical outbursts in the long time span of observations. Observations are taken during the darkest 21 nights per 28 days, in sets of four images, separated by 10 min. Exposures typically last for 30 seconds. Almost 2000 CVs have been detected since 2004 using the three telescopes carrying out this survey, MLS a 1.5 m telescope has detected 1214 CVs/SN, CSS a 0.7 m telescope has detected 601/SN and SSS a 0.5 m telescope has detected 109 CVs/SN. Many of these are new CVs and are detected out to $\sim$ 4 kpc \cite{drak14}. Many of these systems are among the shortest period CVs, demonstrating that deeper limiting magnitudes probe a different population of systems compared to the known bright CVs. \cite{drak14}.

The All-Sky Automated Survey for Supernovae (ASAS-SN) network currently consists of 24 telescopes, across the world and therefore surveys the whole visible sky. Aside from supernovae, it discovers many other transients \cite{jaya18}, including accreting white dwarfs, where almost 3000 CVs and symbiotic systems have been detected, where machine learning methods are used intensively to classify objects \cite{jaya19}. Work on individual subgroups of CVs has also been prolific, such as the analysis of superhump maxima for 56 SU UMa-type dwarf novae \cite{kato14}, where these data support the idea that disk tilt is responsible for modulating the outburst pattern in SU UMa-type dwarf novae.

\subsection{Accreting white dwarfs found in X-ray surveys}

X-ray surveys can reveal different types of accreting white dwarfs compared to the optical surveys. One advantage of the X-ray surveys is that many fewer objects emit at such high energies, making it easy to pick out the accreting white dwarfs. However, the X-ray/$\gamma$-ray data can be ambiguous and follow-up observations are often important to investigate the nature of the X-ray source.

As noted in Section~\ref{sec:polsandips}, magnetic CVs are particularly bright in X-rays  and many of them have been discovered in this domain e.g. \cite{muka17,webb18} and references therein. X-ray observations with Rosat (0.1-2.4 keV, 1990-1999) \cite{true82} and later with XMM-Newton (0.2-12.0 keV, since 1999) \cite{jans01} and Chandra (0.2-10.0 keV, since 1999) \cite{weis99} of globular clusters have identified many CV candidates, notably a significant number of magnetic systems e.g. \cite{webb05} and references therein and see Section~\ref{sec:GCs}. INTEGRAL/IBIS has now identified 929 objects of which 79 are CVs \cite{kriv21}. Identifying the hard X-ray sources detected with INTEGRAL/IBIS (15 keV-10 MeV, since 2003) \cite{wink03} and the Swift/BAT (15-150 keV) \cite{gehr04}, \cite{dema20} have shown that about a quarter of the hard Galactic X-ray sources are accreting white dwarfs, including dwarf novae, novae, novalikes, symbiotic systems. They are dominated by intermediate polars (64\% of the 59 INTEGRAL/IBIS accreting white dwarfs and 55\% of the 88 Swift/BAT accreting white dwarfs) and polars (10\% of the INTEGRAL/IBIS accreting white dwarfs and 15\% of the Swift/BAT accreting white dwarfs). They also show that the intermediate polars reach luminosities of 10$^{34}$ erg s$^{-1}$, with a peak at 5 $\times$  10$^{34}$ erg s$^{-1}$ in the 14-159 keV band. These systems also show a soft blackbody (softer for the polars than the intermediate polars), thought to be emanating from the heated polar region.

Thanks to these recent X-ray observations, there is strong evidence that the Galactic ridge X-ray emission may actually be the result of a large population of accreting white dwarfs, and not the result of X-ray illumination of gas clouds, \cite{revn09}. This is in contrast to the Galactic centre X-ray emission, which may be largely attributed to accreting magnetic white dwarfs \cite{hail16}.

The S-CUBED survey, is a $\sim$weekly 60 s survey of 6$^\circ \times 6^\circ$ centred on the Small Magellanic Cloud (SMC) carried out with the Swift X-ray telescope (0.3-10 keV). It has been ongoing since 2016 \cite{kenn18}, and during this time a number of rare Be star-white dwarf binaries (see Section~\ref{sec:otherCVs}) have been discovered, e.g. \cite{coe20}. The binaries were found as they brightened during an outburst, in which they reach a few percent of the Eddington luminosity. Only 4 such objects are known, even though a large population is predicted \cite{ragu01}, whereas more than 100 X-ray binaries have been found. Using such a survey may reveal many more of these rare objects \cite{kenn21}.

Repeated surveys of large areas with soft X-ray telescopes such as XMM-Newton, Chandra or Swift/XRT (0.3-10.0 keV, since 2004) are the ideal way to discover Super soft sources, see Section~\ref{sec:CN_sss}. Less than 20 such sources are known, with many in the Large Magellanic Cloud (LMC) discovered with the Einstein observatory \cite{long81} (0.15-4.5 keV, 1978-1981) or Rosat. They can be difficult to disentangle from magnetic CVs in our own galaxies, and indeed in a recent survey of the LMC a further four were discovered, although two maybe magnetic CVs in the Milky Way \cite{maitr21}. The large number of supersoft systems in the Magellanic clouds may simply be thanks to the large area surveys and the low column density which allows these sources with very soft spectra to be detected. In a study of supersoft sources in 4 spiral galaxies, 2 lenticular galaxies and 3 ellipticals \cite{gali21} showed that there is about a factor 8 less SSS in early-type galaxies compared to later type galaxies, in alignment with results from population synthesis codes \cite{gali21}.

\subsection{Future surveys that will detect accreting white dwarfs}
\label{sec:future_survey}

Amongst the future missions that will be interesting for studying accreting white dwarfs, four are highlighted in this section. Although the X-ray detector eRosita (0.2-10.0 keV)  on board the observatory Spektrum-RG was launched in 2019 \cite{pred21}, the data will not be released for a few years, so the results are yet to come. It will survey the whole sky several times in its lifetime. An early release of a portion of the sky was made recently, giving insight into the full data release that will occur in a few years time. A number of interesting accreting white dwarfs have already been published, e.g. SRGt 062340.2$-$265715, a new novalike CV at a distance of 495 pc, possibly an X-ray underluminous magnetic CV, an intermediate polar, or an overluminous nonmagnetic CV \cite{schwo21}, but many more accreting white dwarfs are expected. \cite{schw12} stated that eRosita should reach a flux limit of about 10$^{-14}$ erg cm$^{-2}$ s$^{-1}$ (0.5$-$2 keV) so it may be possible to discover all CVs brighter than log L$_X$ = 29, 30, 31, 32 erg s$^{-1}$ out to 0.3, 0.9, 3, and 9 kpc, respectively.  \cite{schw12} estimated that as many as 9000 CVs will be detected in half of the sky or 18000 in the whole sky.

The Vera C. Rubin observatory \cite{ivez19} will start to observe the sky systematically and repeatedly from early 2024. Using the bands $u, g, r, i, z$ and $y$, it will reach magnitudes over the 10 years as deep as 27.5 for the $r$ band and as deep as 25th magnitude in $g$ for a single visit. The high cadence observations will allow thousands of new accreting white dwarfs to be discovered across the galaxy and the frequent observations will allow orbital periods to be determined and outbursts to be studied. 2000 double degenerates are also expected to be discovered through the identification of periodicities \cite{koro17} and studying their evolution will provide an insight into the type Ia supernovae. 

In the 2030s, two major missions from the European Space Agency will be launched, the Laser Interferometer Space Antenna (LISA) \cite{amar17} that will detect gravitational waves in the frequency range from tens of micro-hertz to 1 Hz and the Advanced Telescope for High Energy Astrophysics (Athena) \cite{nand13} that will observe the sky with a sensitive Silicon Pore Optics telescope in the 0.2-12 keV range. The most prevalent sources detected with LISA will be Galactic double white dwarfs that are about a hundred thousand years before merger e.g. \cite{lamb19}. Using cosmological simulations of our galaxy along with binary population synthesis models \cite{lamb19} suggest that as many as 12000 double white dwarfs could be detected with LISA. These include Helium-Helium systems, Helium-Carbon/Oxygen systems and Carbon/oxygen-Carbon/oxygen systems. Athena will be able to  provide unique plasma diagnostics of the disk-white dwarf boundary layer in CVs, and the magnetically-channelled accretion flows in (intermediate) polars using the X-ray Integral Field Unit (X-IFU). Athena will be sensitive enough to routinely measure the gravitational red-shifted iron K$_\alpha$ line from the surface of the white dwarf which will give a constraint on the mass of the white dwarf\cite{nand13}. Athena
observations will measure the temperature, mass, and chemical composition of novae and supersoft sources to get insight into SN Ia progenitors \cite{motc13}.

\section{Open questions regarding accreting white dwarfs}

A number of different questions have been addressed in this Chapter and whilst some of the questions posed have been partially answered, progress is required to fully understand these problems. One of the most intriguing questions is why are the white dwarf masses in accreting white dwarfs on average more massive than those in the field? To determine this, the evolution of these systems should be understood, notably the quantity of angular momentum lost. It will also be necessary to understand the physics behind the mass increase, be it through accretion, mergers or something else. This could also help elucidate the low space density of CVs observed, as well as the number of CVs observed above and below the period gap and the number of period bouncers observed, compared to theoretical predictions. It may also give insight into the mechanism for type 1a supernovae, which still needs to be understood.

Another area which requires a deeper understanding is the radio emission associated with jet emission from these systems. Are the CV jets equivalent to jets observed in other accreting systems? What is the physical process behind the jet emission and why do some accreting white dwarfs not show jets?

A number of questions remain around the CVs in globular clusters, namely what is the origin of these systems, are the centrally located systems really the products of encounters and those located in the periphery evolved from primordial binaries? Also, why are so few outbursts observed from these CVs in globular clusters compared to those in the field? 

Future observations will give further insight into the accretion processes in accreting white dwarfs. Whilst this has not been discussed in detail in this Chapter, it is still a subject of intense research as it will also help in understanding the evolution of these systems and other accreting systems such as supermassive black holes in the centres of galaxies.  Upcoming surveys, as described in the previous sections, will also provide a wealth of new accreting white dwarfs and insight into these systems which will allow many of the questions discussed here to be resolved in the next two decades.

\section{Cross-References}

Chapter~\ref{ch:acc_disk_theory} : Overall accretion disk theory, Shane Davis \\
Chapter~\ref{ch:binary_evol} : Overall binary evolution theory, Diogo Belloni \\
Chapter~\ref{ch:white_dwarfs} : White dwarfs, Kim Page \\
Chapter~\ref{ch:accretion} : Accretion process, Thomas J. Maccarone \\
Chapter~\ref{ch:ejection} : The extreme of SMBH accretion/ejection, William Alston \\

   \bibliographystyle{spbasic} 
   \bibliography{WebbV2} 

%
%
%
%
%

\end{document}